\documentclass[aps,prl,groupedaddress,twocolumn,10pt,reprint,superscriptaddress, longbibliography,floatfix]{revtex4-2}

\usepackage{amsmath}
\usepackage{amssymb}
\usepackage{amsthm}
\usepackage{bbm}
\usepackage{graphicx}
\usepackage{amsmath}
\usepackage{graphicx}
\usepackage{color}
\usepackage{upgreek}
\usepackage[linkcolor = blue, citecolor = blue, urlcolor = blue, colorlinks = true]{hyperref}
\usepackage{bm}
\usepackage[capitalise]{cleveref}
\usepackage{textcomp}
\usepackage{hyperref}
\usepackage[usenames,dvipsnames]{xcolor}
\usepackage[normalem]{ulem}
\usepackage{wasysym,siunitx}
\usepackage{comment}

\usepackage{siunitx}

\hypersetup{colorlinks=true, linkcolor=blue!50!black, urlcolor=blue!50!black, citecolor=blue!50!black}

\begin{document}

\title{Ultrasensitivity without conformational spread: A mechanical origin for non-equilibrium cooperativity in the bacterial flagellar motor} 

\author{Henry H. Mattingly}
\email{hmattingly@flatironinstitute.org}
\affiliation{Center for Computational Biology, Flatiron Institute, New York, NY, USA}
\author{Yuhai Tu}
\email{yuhai@us.ibm.com}
\affiliation{IBM Thomas J. Watson Research Center, Yorktown Heights, NY, USA}

\date{\today}

\begin{abstract}

    Flagellar motors enable bacteria to navigate their environments by switching rotation direction in response to external cues with high sensitivity. Previous work suggested that ultrasensitivity of the flagellar motor originates from conformational spread, in which subunits of the switching complex are strongly coupled to their neighbors as in an equilibrium Ising model. However, dynamic single-motor measurements indicated that rotation switching is driven out of equilibrium, and the mechanism for this dissipative driving remains unknown.     Here, based on recent cryo-EM structures, we propose that local mechanical torques on motor subunits can affect their conformation dynamics. This gives rise to a tug of war between stator-associated subunits, which produces cooperative, non-equilibrium switching responses without requiring nearest-neighbor interactions. Since subunits are effectively coupled at a distance, we call this mechanism ``Global Mechanical Coupling." Our model makes a qualitatively new prediction that the motor response cooperativity grows with the number of stators driving rotation. Re-analyzing published motor dose-response curves in varying load conditions, we find tentative experimental evidence for this prediction. Finally, we show that operating out of equilibrium enables motors to achieve high cooperativity with faster responses compared to equilibrium motors. Our results suggest a general role for mechanics in sensitive chemical regulation. 
\end{abstract}

\maketitle


The bacterial flagellar motor (BFM) is a complex, macromolecular machine \cite{armitage_assembly_2020, wadhwa_bacterial_2021, guo_bacterial_2022, hu_structural_2022} that propels motile bacteria. In response to external cues, the BFM can switch its direction of rotation, e.g. from counter-clockwise (CCW) to clockwise (CW), which reorients the cell body and enables directed navigation \cite{berg_chemotaxis_1972}.
Different species exhibit diverse motor architectures and swimming patterns \cite{grognot_more_2021}, yet many of the core components are conserved \cite{chen_structural_2011, zhao_molecular_2014, carroll_structural_2020}. 

While much is known about the BFM, fundamental questions about the rotation switching decision remain. In \textit{Escherichia coli}, switching is extremely sensitive to the internal concentration of the response regulator CheYp, with a Hill coefficient of at least 10 \cite{cluzel_ultrasensitive_2000, yuan_ultrasensitivity_2013}. Based on the observation that the switch complex contains a ring of $\sim 34$ FliM subunits to which CheYp binds \cite{thomas_rotational_1999}, Duke et al. \cite{duke_conformational_2001} proposed a ``conformational spread" mechanism wherein each FliM subunit can be in one of two conformational states (CW or CCW), and neighboring FliM's prefer to be in the same conformation. Quantitatively, the conformational spread mechanism can be described by an equilibrium Ising-type model with strong nearest-neighbor coupling, which leads to a high Hill coefficient in agreement with experiments \cite{duke_conformational_2001, bai_conformational_2010}.

However, measurements of single motor switching dynamics have found that the distributions of CCW and CW intervals exhibit a peak, first observed by Korobkova et al. \cite{korobkova_hidden_2006} and later confirmed by Wang et al. \cite{wang_non-equilibrium_2017} (see also \cite{tu_driven_2017}). Theoretically, it was shown by Tu \cite{tu_nonequilibrium_2008} that peaked interval-time distributions cannot be generated by any equilibrium process. This general result implies that the BFM switching dynamics must be coupled to energy dissipation, mostly likely the ion motive force which drives rotation of the flagellum. Subsequent theoretical work has added non-equilibrium effects \textit{ad hoc} to the conformational spread model \cite{wang_non-equilibrium_2017, wang_dynamics_2021, zhu_mechanosensitive_2024}, but the physical mechanism by which rotation switching is coupled to dissipation remains unclear. 

Recently, key insights about the rotation switching mechanism emerged from cryo-EM structures of BFMs \cite{deme_structures_2020, santiveri_structure_2020}. First, the MotAB stators that drive motor rotation in many bacterial species are composed of 5 MotA and 2 MotB monomers, and they themselves rotate. Torque is transduced to the flagellar rotor through electrostatic interactions between MotA and FliG in the C-ring at the base of the flagellum. Thus, the C-ring can be thought of as a large gear that is rotated by smaller stator gears, where MotA and FliG are the cogs \cite{johnson_structural_2024}. Structures of motors locked in different rotation states revealed that FliG subunits interact with the inner edge of the stators during CCW rotation, but change pose by 20-30 degrees to interact with the outer edge during CW rotation (\cite{chang_molecular_2020, carroll_flagellar_2020}, Fig. \ref{fig:schematic}A). This suggests that stators always rotate CW, and conformation changes in FliG cause the flagellum to switch rotation direction. Binding of CheYp to FliM then biases its associated FliG towards the outer conformation to induce CW rotation.

Based on these structures, we propose a model for how the internal mechanics of the motor could coordinate FliG subunits and generate a non-equilibrium, cooperative switching response without requiring conformational spread. The central idea is that, when the conformation of one FliG subunit differs from the rest, the local torque from its stator could mechanically push it into alignment with the others. Since forces transduced through the C-ring effectively couple all stator-engaged FliG subunits, we call this ``Global Mechanical Coupling" (GMC). Using a coarse-grained theory and simulations, we show that GMC dissipates energy to generate high cooperativity. Furthermore, GMC uniquely predicts that the Hill coefficient of the switching response increases with the number of stators driving rotation, and we provide tentative evidence for this prediction. Finally, we show that GMC works synergistically with the nearest-neighbor couplings of conformational spread, easing a speed-sensitivity trade-off faced by purely equilibrium models. The GMC mechanism provides an unique starting point for understanding how chemical and mechanical inputs are integrated into motor function.

\section{Results}

\subsection{Local torques can drive FliG conformation cycles} 

\begin{figure}[]
	\includegraphics[]{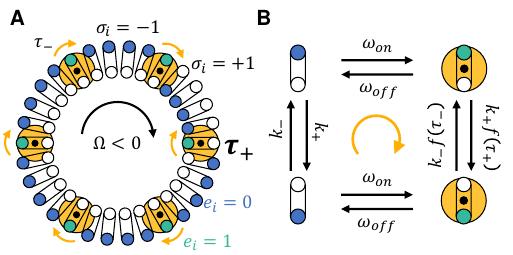}
	\caption{A) Schematic top view of the motor, showing FliG subunits of the C-ring (blue, green circles) and stators (yellow circles) (see e.g. \cite{wadhwa_bacterial_2021, chang_structural_2021, guo_bacterial_2022, hu_structural_2022}). At a moment in time, FliG subunits are either in an inner conformation ($\sigma_i=+1$) or outer conformation ($\sigma_i=-1$), and they can be engaged with a stator (green, $e_i=1$) or unengaged (blue, $e_i=0$). FliG subunits translate through space, engaging and unengaging with stators, as the C-ring rotates at angular velocity $\Omega$.
	Stators are fixed in place and exert torques in the directions of the yellow arrows. When a stator-engaged FliG subunit is unaligned with the other engaged subunits, C-ring rotation drives it against the torque of its stator. Thus, this subunit experiences a larger torque (here, $\tau_+$) than those aligned with the rotation direction (here, $\tau_-$). B) Local torques can drive FliG conformation dynamics out of equilibrium.}
	\label{fig:schematic}
\end{figure}

What happens when one FliG subunit adopts a conformation that is different from the rest? As the C-ring rotates, that subunit will encounter a stator and be pushed \textit{against the stator's torque} (Fig. \ref{fig:schematic}A).
Thus, the unaligned subunit should experience a larger torque than those aligned with the direction of rotation. If this local torque mechanically pushes the subunit towards the opposite conformation, this could coordinate subunit conformation dynamics. 

To make this picture mathematical, we construct a minimal model of FliG-stator mechanics and FliG conformation dynamics. We model the C-ring as a ring of $N$ FliG subunits, where each site is characterized by a binary conformation, $\sigma \in \{-1,+1\}$, and a binary stator-engagement state, $e\in\{0,1\}$ (Fig. \ref{fig:schematic}AB). 
Here, $\sigma=+1$ and $-1$ correspond to the inner and outer conformations of FliG subunits observed in cryo-EM experiments. Then, $e=1$ when a subunit is in contact with a stator, and $e=0$ otherwise. The number of stators driving the flagellum is $M=\sum_{i=1}^N \, e_i(t)$. $M$ is known to adapt to external load on a time scale of minutes \cite{blair_restoration_1988, lele_dynamics_2013, tipping_load-dependent_2013, tusk_subunit_2018,  wadhwa_torque-dependent_2019, nirody_load-dependent_2019, wadhwa_mechanosensitive_2021, wadhwa_multi-state_2022}, up to $M\approx11$ \cite{reid_maximum_2006}, but we will consider constant-load environments and take $M$ to be constant.

Subunits that are not engaged with a stator, $e=0$, switch conformation from $\sigma=+1\rightarrow-1$ and vice versa at rates $k_+$ and $k_-$:
\begin{equation}\label{eq:k}
	k_{\pm} = k_0 \exp(\mp \Delta F/2).
\end{equation}

\noindent The free energy difference, $\Delta F$, between conformations $\sigma=+1$ and $-1$ includes the effects of CheYp binding and intrinsic energy differences, and $k_0$ is the switching rate when $\Delta F = 0$. Throughout, we rescale energy by the scale of thermal fluctuations, $k_B\, T \approx 4.1$ pN nm at room temperature.

When engaged with a stator, the switching rates are modified by the torque, $\tau$, exerted by the stator on the subunit:
\begin{equation}\label{eq:ke}
	k_+^e = k_+ \, f(\tau_+), \quad k_-^e = k_- \, f(\tau_-)
\end{equation}

\noindent where $\tau_\pm$ is the torque on subunits in conformation $\sigma = \pm1$, and $f(\tau)$ is a monotonically increasing function. Here, we take:
\begin{equation}\label{eq:f}
	f(\tau) \equiv \exp(\gamma \, \tau).
\end{equation}
	
\noindent This functional form is commonly used to model slip-bonds \cite{bell_models_1978, wiita_force-dependent_2006}, where protein-protein or protein-ligand dissociation rates increase with force along a reaction coordinate. With local torque $\tau$ rescaled by the stator stall torque $\tau_0$ \cite{ryu_torque-generating_2000, yuan_resurrection_2008, yuan_switching_2009, nakamura_effect_2009}, the dimensionless parameter $\gamma \approx \alpha \, \tau_0 \, \Delta \phi / (k_B \, T)$ includes: the projection of the stator torque along the direction of the FliG conformation transition, $\alpha$; the difference in pose angles of the two FliG conformations, $\Delta \phi \approx 0.4$ radians \cite{chang_molecular_2020, carroll_flagellar_2020}; and the scale of thermal fluctuations. Below, we will vary the value of $\gamma$, but plugging in numbers, with $\alpha \approx 0.1$, we get $\gamma \approx 3.2$.
		
The conformation switching dynamics of this model dissipate energy. In the mean field limit, given the states of the rest of the subunits, a single FliG subunit can be in one of four states (Fig. \ref{fig:schematic}B). In addition to conformation switching dynamics, subunits transition between stator-engaged and -unengaged states with rates that depend on the motor rotation speed $\omega_{on/off} \propto \Omega$. With four states, a subunit can undergo cycles in its phase space, and the free energy dissipated per cycle is:
\begin{equation}\label{eq:G_cycle}
	\Delta G = \ln\left(\frac{\omega_{on}\,k_+\,f(\tau_+)\,\omega_{off}\,k_-}{k_+ \, \omega_{on}\, k_- \, f(\tau_-)\, \omega_{off}}\right)= \gamma \, (\tau_- - \tau_+),
\end{equation}

\noindent using $f(\tau)$ above. Thus, detailed balance is broken as long as $\gamma\ne 0$ and $\tau_-\neq\tau_+$. Microscopically, $\tau_-\neq\tau_+$ because torque is distributed evenly among subunits, so those in the majority conformation experience smaller torques per subunit. Thus, rotation switching can be driven out of equilibrium by torque-dependent conformation dynamics. Rather than ATP hydrolysis, the energy source for this driving is the ion motive force across the cell membrane, which is necessary for the stators to apply torques on FliG and the C-ring.

\subsection{Minimal model of motor mechanics}

Next, we need a model for the local torque on stator-engaged FliG subunits. We assume subunits have strong interactions with stators, so that the two rotate together like gears. Furthermore, we assume that the rotor rotates as a rigid object. We also assume that the system is in the over-damped limit, that stator refueling and release of ions are fast, and we neglect fluctuations in torque. Below, we rescale time by the motor's zero-load speed, $\Omega_0$, and torque by the stator stall torque, $\tau_0$, which in \textit{E. coli} are $\Omega_0 \approx 300$ Hz and $\tau_0 \approx 300$ pN nm \cite{ryu_torque-generating_2000, yuan_resurrection_2008, yuan_switching_2009, nakamura_effect_2009}.

With these assumptions, we write down torque balances on each stator-engaged FliG subunit and on the rotor, along with no-slip conditions between FliG's and stators (SI). Solving these equations for the dimensionless angular velocity of the motor, $\Omega(t)$, gives:
\begin{equation}\label{eq:Omega}
	\Omega(t) = \frac{M/\beta}{1+M/\beta} \left(2 \frac{N_e^+(t)}{M} - 1 \right),
\end{equation}

\noindent where $\beta$ is the dimensionless rotational drag on the flagellum and $N_e^+(t) = \sum_{i | \sigma_i(t)=+1} e_i(t)$ is the number of stator-engaged FliG subunits in the $\sigma=+1$ conformation. Thus, the rotation velocity depends linearly on the number of engaged subunits in the $+1$ conformation, and the rotation direction is set by the majority conformation among engaged subunits. At full alignment ($N_e^+=0$ or $N_e^+=M$), the rotation speed  depends on $M/\beta$, which can be understood as the ratio of the motor's output stall torque relative to the drag torque at the zero-load rotation speed (SI). From recent experiments in \textit{E. coli} \cite{niu_flagellar_2023}, $M \approx 4-6$ and $\beta \approx 2.5$ during swimming. 

The torque balances also give us the magnitude of the local torque on stator-engaged FliG subunits:
\begin{equation}\label{eq:tau}
	\tau_\pm(t) = 1 \mp \Omega(t).
\end{equation}

\noindent As previewed earlier, $\tau_-\neq\tau_+$ when the motor is rotating. Together with Eqn. \eqref{eq:f}, the expression in Eqn. \eqref{eq:tau} implies that local torques change both the free energy difference and the energetic barrier between conformations $\sigma=+1$ and $-1$.

Since $\Omega(t)$ depends on the states of \textit{all} stator-engaged subunits, local torques on subunits depend on the global state of the motor. Eqns. \eqref{eq:Omega} and \eqref{eq:tau} indicate that FliG subunits that are in the majority conformation experience smaller torques than those in the minority, and the difference between them increases as the size of the majority increases. Furthermore, the local torque on stators in the minority exceeds their stall torque ($\tau_\pm>1$). With the no-slip condition, this predicts that stators in the minority can be overpowered by the majority and forced to rotate CCW, \textit{against} their natural rotation direction (SI Fig. 1).

\subsection{Global mechanical coupling coordinates subunits}

The switching rates and local torques above indicate that the stator-engaged subunits are in a tug of war. Say the motor is initially not rotating, $\Omega(t)=0$, and therefore half of the engaged subunits are in the conformation $\sigma=+1$, $N_e^+=M/2$. When one subunit flips conformation by chance, the tie is broken and the motor begins to rotate. Then, subunits in the minority conformation experience a larger torque than those in the majority (Eqn. \eqref{eq:tau}), increasing the rate at which they flip into the majority (Eqns. \eqref{eq:ke} and \eqref{eq:f}). This creates a positive feedback loop that rapidly pushes all subunits into alignment.

To study this model quantitatively, we first considered a coarse-grained theory. Motivated by the fact that the motor rotation velocity and local torques only depend on $N_e^+$, we consider the dynamics of just two variables: the number of engaged, $N_e^+(t)$, and unengaged, $N_u^+(t)$, FliG subunits in the $\sigma=+1$ conformation. Conservation of stators determines $N_e^-(t)$, $N_e^+(t) + N_e^-(t) = M$, and conservation of FliG subunits determines $N_u^-(t)$, $N_u^+(t) + N_u^-(t) = N-M$, where $N_e^-$ and $N_u^-$ the number of engaged or unengaged subunits in the $\sigma=-1$ conformation. While this coarse-graining fully captures the conformation switching dynamics, exchange of stator-engaged FliG subunits depends on the full ring structure and must be approximated (SI). To reduce the number of model parameters, we will take $M/\beta\gg1$.

\begin{figure}[b]
	\includegraphics[]{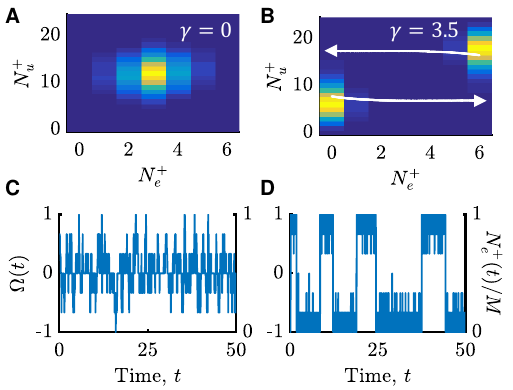}
	\caption{Microscopic Gillespie simulations of the full FliG ring, projected into the subspace $(N_e^+,N_u+)$. Throughout, we use $N=30$, $k_0=1$, $M = 6$, $M/\beta\gg1$, and $\Delta F = 0$, unless specified otherwise. A,B) Stationary distributions of $(N_e^+,N_u^+)$, with probability increasing from dark to light. A) Independent switching of FliG subunits ($\gamma=0$). B) Increasing $\gamma$ leads to a bimodal distribution. White arrows show average transition paths from state $N_e^+=0$ to $N_e^+=M$, and vice versa. C,D) Time series of motor rotation velocity $\Omega$ or $N_e^+/M$ (Eqn. \eqref{eq:Omega}). C) When $\gamma=0$ (as in A), rotation velocity fluctuates around zero. D) At larger $\gamma$ (as in B), the motor shows extended bouts of rotation at nearly full speed in either direction.}
	\label{fig:bimodal}
\end{figure}

In the coarse-grained model, we can derive the stationary distribution of $N_e^+$, and thus of $\Omega$. When $k_0\gg1$, we find:
\begin{equation}\label{eq:PNe}
	P(N_e^+) = \frac{1}{Z} \binom{M}{N_e^+} \exp\left( \Delta F \, N_e^+ + 2\,\gamma \, M \left( \frac{N_e^+}{M} - \frac{1}{2} \right)^2 \right)
\end{equation}

\noindent with normalization constant $Z$. The stationary distribution $P(N_e^+)$, like any one-dimensional projection, has the form of a Boltzmann distribution \cite{gardiner_stochastic_2009}. The binomial prefactor is an entropic term that favors a monomodal distribution centered on $N_e^+=M/2$, the no-rotation state. In the exponential is an effective Hamiltonian that has contributions from the bias, $\Delta F$, and a term of non-equilibrium origin that is equivalent to a sum of effective energetic couplings among all pairs of stator-engaged subunits (SI). The latter, positive-feedback term favors macrostates with consensus among engaged subunits, $N_e^+=0$ or $N_e^+=M$. As $\gamma$ increases, this term leads to a bimodal distribution of $N_e^+$. In the limit of large $\gamma$, eventually all weight is on the two consensus macrostates, resulting in a non-equilibrium Monod-Wyman-Changeux (MWC) model \cite{monod_nature_1965} for motor rotation (SI Eqn. 38). Thus, the internal mechanics effectively couple all stator-engaged FliG subunits, hence the name ``Global Mechanical Coupling" (GMC). 

Gillespie simulations \cite{gillespie_stochastic_2007} of the ``microscopic" ring structure, projected onto the subspace $(N_e^+, \, N_u^+)$, agreed with these predictions and provided motor switching dynamics (Methods, Fig. \ref{fig:bimodal}). When there was no mechanical coupling, $\gamma=0$, the time series of motor state showed binomial fluctuations around $N_e^+=M/2$ (Fig. \ref{fig:bimodal}AC). At higher values of $\gamma$, the stationary distribution of $(N_e^+, \, N_u^+)$ became bimodal (Fig. \ref{fig:bimodal}B). Furthermore, the time series of motor state showed extended bouts of rotation at nearly full speed in either direction, $\Omega(t) \approx \pm1$, with fast transitions (Fig. \ref{fig:bimodal}D), similar to experimental measurements (e.g. \cite{bai_conformational_2010}). Furthermore, switching of rotation direction from CW to CCW and vice versa took different transition paths (Fig. \ref{fig:bimodal}B; SI; \cite{zakine_minimum-action_2023, hathcock_time-reversal_2024}), a clear indication of broken detailed balance.

\subsection{GMC produces high response cooperativity at the cost of free energy dissipation}

A striking feature of the BFM is the high cooperativity of its responses to intracellular CheYp concentration. To quantify cooperativity in the GMC model, we computed the Hill coefficient, $H$, of the stationary mean $\langle N_e^+\rangle/M$ (and thus of rotation velocity $\langle \Omega \rangle$) as a function of the free energy bias, $\Delta F$ \cite{hill_possible_1910, duke_conformational_2001}.

In the coarse-grained theory:
\begin{equation}\label{eq:Hill}
	H \approx \frac{4}{M} \, \partial_{\Delta F} \langle  N_e^+ \rangle \big |_{\Delta F = 0} \approx \frac{4}{M} Var(N_e^+) \big |_{\Delta F = 0} \leq M.
\end{equation}

\begin{figure}[t]
	\includegraphics[]{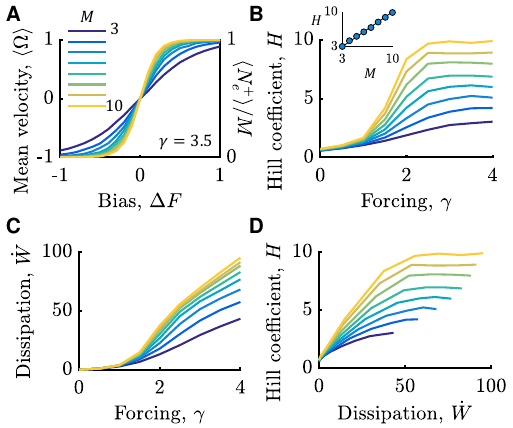}
	\caption{Non-equilibrium motor cooperativity. A) Stationary mean rotation velocity, $\langle \Omega \rangle$, or $\langle N_e^+\rangle/M$, versus free energy bias, $\Delta F$. Responses for varying $M$ are shown, increasing from 3 (light) to 10 (dark), with $\gamma=3.5$ fixed. Colors are the same in all panels. B) Hill coefficients $H$ as a function of $\gamma$, for various values of $M$. Inset is $H$ versus $M$ at $\gamma=4$, where markers are from simulation data and the solid line is $y=x$. C) Free energy dissipation rate versus $\gamma$, for various values of $M$. D) Hill coefficient (B) versus dissipation rate (C).}
	\label{fig:Hill}
\end{figure}

\noindent This predicts that the Hill coefficient is proportional to the variance of $N_e^+$, which increases with $\gamma$ as the distribution becomes bimodal. In the large-$\gamma$ limit, the Hill coefficient approaches $M$, the number of stators. Thus, GMC makes a qualitatively new prediction: the cooperativity of the motor switch to \textit{chemical} inputs increases with the number of \textit{torque generators} driving rotation. Microscopic Gillespie simulations were consistent with this prediction (Fig. \ref{fig:Hill}AB). In SI Fig. 2, we verify that this prediction persists even when there is nonzero nearest-neighbor coupling between subunits, and that it requires mechanical coupling, $\gamma>0$.

While increasing $\gamma$ increased the cooperativity, $H$, it also increased the free energy dissipation rate. We computed the energy cost of cooperativity, $\dot{W}$, from the Kullback-Leibler divergence between forward and backward simulation trajectories as a function of $\gamma$ (Fig. \ref{fig:Hill}C; Methods; \cite{kullback_information_1951, kawai_dissipation_2007, yu_energy_2022}). Plotting $\dot{W}$ versus $H$ in Fig. \ref{fig:Hill}D showed that increasing dissipation ``buys" steeper responses, up to the point where $H$ saturates at $M$.

Recent experiments provide tentative evidence in support of the prediction that the Hill coefficient increases with the number of stators driving rotation. Zhu et al. \cite{zhu_mechanosensitive_2024} measured the CW bias of individual motors in multiple load conditions by varying the concentration of Ficoll in the experimental medium. Since stator number $M$ increases with load due to adaptation \cite{lele_dynamics_2013, tipping_load-dependent_2013, wadhwa_torque-dependent_2019, wadhwa_mechanosensitive_2021, wadhwa_multi-state_2022}, a corresponding increase in the Hill coefficient would predict specific changes in motor CW bias, assuming the average CheYp concentration in each cell is unchanged. 
	
In particular, we used measurements of CW bias from 40 motors monitored by 0.35-$\mathrm{\mu m}$ beads in $0\%$ and $19\%$ w/v Ficoll (Fig. 3 of \cite{zhu_mechanosensitive_2024}). In the former condition, about half of the stator sites are occupied ($M\approx6$), while in the latter nearly all are occupied ($M\approx 11$) \cite{tipping_load-dependent_2013, zhu_mechanosensitive_2024}. We fit these data to estimate each cell's internal CheYp concentration, as well as the values of the load-dependent Hill coefficient, $H$, and half-maximum CheYp concentration, $K$ (details in the SI). To uniquely specify parameter values, we fixed $H(19\%) = 10.3$ and $K(19\%) = 3.1 \, \mathrm{\mu M}$, as in Ref. \cite{zhu_mechanosensitive_2024}.

\begin{figure}[t]
	\includegraphics[]{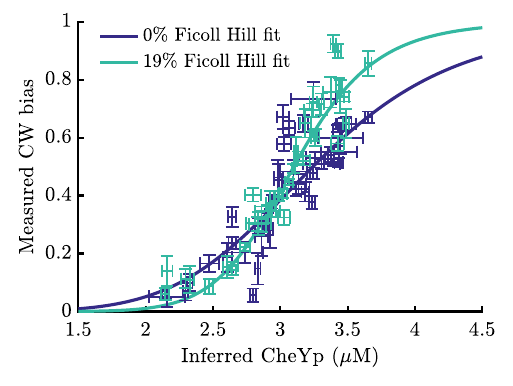}
	\caption{Tentative evidence for GMC. Measured CW bias versus inferred CheYp concentration for individual motors in two load conditions, $0\%$ Ficoll and $19\%$ Ficoll, along with Hill function fits. Error bars are standard errors, where vertical bars are measurement uncertainties and horizontal bars are uncertainties from the inference procedure. Data are from Ref. \cite{zhu_mechanosensitive_2024}.}
	\label{fig:data}
\end{figure}

Fig. \ref{fig:data} shows the measured CW biases as a function of inferred CheYp concentration, along with the Hill function fits. The best-fit parameter values were $H(0\%) = 6.0 \pm 0.3$ and $K(0\%) = 3.24 \pm 0.01 \, \mathrm{\mu M}$, suggesting that the Hill coefficient nearly doubles upon an increase in load from $0\%$ to $19\%$ Ficoll. Since previous measurements suggest that the adapted $M$ roughly doubles across these loads \cite{tipping_load-dependent_2013}, and GMC predicts $H \propto M$ under certain conditions (Eqn. \eqref{eq:Hill}), these results are very suggestive. Simultaneous measurements of $M$, CheYp, and CW bias are needed to fully test this prediction.
	
\subsection{GMC eases a speed-sensitivity trade-off in equilibrium models} 

Conformational spread can reproduce the motor's highly cooperative response to chemical inputs without dissipating energy \cite{duke_conformational_2001, bai_conformational_2010}. Furthermore, while the Hill coefficient $H$ of GMC alone is bounded by the number of stators when $k_0\gg1$ ($H\leq M$), nearest-neighbor interactions can achieve $H$ up to the size of the entire C-ring ($H\leq N$) \cite{owen_size_2023}. Why might the BFM dissipate energy in the switching response? Detailed balance constrains the dynamics of equilibrium models, and past work has shown that energy dissipation can generally ease trade-offs between speed and sensitivity or accuracy \cite{lan_energyspeedaccuracy_2012, sartori_free_2015, fei_design_2018, hathcock_time-reversal_2024}.

To test this possibility, we computed the Hill coefficient and the response speed, $\tau_R^{-1}$, for varying values of $\gamma$ and nearest-neighbor coupling, $J$. Response speed was defined as the average time it took a motor in the stationary state with $\Delta F=1/2$ to reach $N_e^+=0$ after a rapid change in bias to $\Delta F=-1/2$, mimicking a tumble after a change in swimming direction. This response speed depends on the values of $\Delta F$ before and after the ``stimulus," but we expect similar qualitative behavior regardless of the specific values.

\begin{figure}[t]
	\includegraphics[]{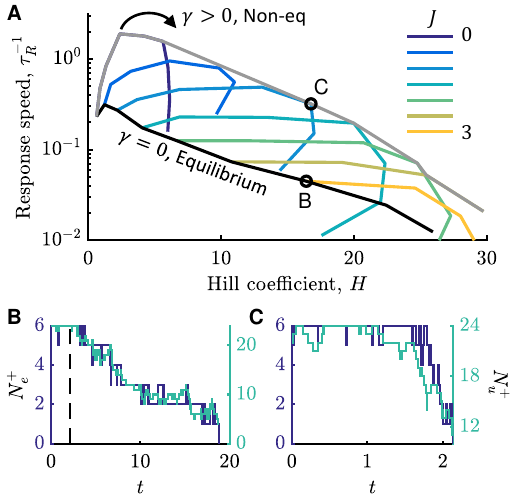}
	\caption{GMC eases speed-sensitivity trade-off of equilibrium motors. A) Response speed, $\tau_R^{-1}$, versus Hill coefficient, $H$, for equilibrium motors with nearest-neighbor interactions (black line, $\gamma=0$, $J>0$) and for non-equilibrium motors (colorful lines, $\gamma>0$, $J\geq0$). In the latter, each color is a fixed value of $J$ with $\gamma$ increasing clockwise around the curve. Gray is the envelope of optimal strategies. B,C) Representative response trajectories of equilibrium (B) and non-equilibrium (C) motors marked in (A), which have similar $H$ but 10-fold different $\tau_R^{-1}$. Plotted are time series of $N_e^+$ (blue, left axis) and $N_u^+$ (teal, right axis). Vertical dashed line in (B) is the duration of (C). In (B), $J=3$, $\gamma=0$. In (C), $J=1$, $\gamma=2$.}
	\label{fig:synergy}
\end{figure}

As $J$ increased, equilibrium conformational spread models ($\gamma=0$) showed decreasing response speed with increasing Hill coefficient (Fig. \ref{fig:synergy}A, black line). Making $\gamma$ non-zero introduced the non-equilibrium effects of GMC. Initially, increasing $\gamma$ increased \textit{both} $H$ and $\tau_R^{-1}$, up to a point where $H$ peaked and the dynamics slowed down dramatically (Fig. \ref{fig:synergy}A, colorful lines). Non-equilibrium motors with both $J$ and $\gamma$ non-zero consistently achieved the same Hill coefficient as equilibrium ones with about 10 times faster responses.

Representative trajectories demonstrate why GMC enables faster responses. In Fig. \ref{fig:synergy}BC, we compare an equilibrium motor and a non-equilibrium one, both of which have a Hill coefficient of $H \approx 16$. The equilibrium motor transitions as described before \cite{duke_conformational_2001}, where a ring of all $\sigma=+1$ subunits nucleates two energetically-unfavorable interfaces with a domain of $\sigma=-1$, which then grows slowly via a drift-diffusion process. Since stator engagement is nearly independent of conformation dynamics, $N_e^+$ and $N_u^+$ change in concert. In the motor with GMC, nearest-neighbor interactions are weaker, so the weakly-coupled, unengaged spins first flip to $\sigma=-1$, causing $N_e^+$ to decrease when those subunits engage with stators. Once $N_e^+$ crosses below $M/2$, positive feedback rapidly pushes $N_e^+$ to zero.

\section{Discussion}

Here, we proposed a biophysical mechanism for the non-equilibrium rotation switching dynamics of the bacterial flagellar motor. In this mechanism, the conformation dynamics of FliG subunits in the C-ring depend on the local torque they experience, which in turn depends on the conformations of all other subunits through rigidity of the C-ring. Since internal mechanics of the motor coordinate subunits at a distance, we call this mechanism ``Global Mechanical Coupling."

Past works have considered torque-dependent FliG conformation dynamics \cite{yuan_switching_2009, bai_coupling_2012, wang_non-equilibrium_2017, wang_dynamics_2021, zhu_mechanosensitive_2024}. One line of work treated the motor \textit{output} torque in a given load condition as a fixed parameter that modulated the conformation dynamics of all FliG subunits \cite{yuan_switching_2009, wang_non-equilibrium_2017, wang_dynamics_2021, zhu_mechanosensitive_2024}. It is unclear how this phenomenological description would be implemented at the molecular level. A separate study considered the effects of instantaneous torque on individual, stator-engaged FliG subunits' conformation dynamics \cite{bai_coupling_2012}. However, it was again assumed that motor output torques influence FliG switching. No previous model of rotation switching considered the equal and opposite torque from the C-ring on the stators, which is required for physical consistency.
Here, a torque balance on the stators and C-ring revealed that local torques on FliG subunits depend on the global motor state.

GMC makes a testable prediction that the Hill coefficient of the motor switching response to chemical inputs, $H$, increases with the number of stators driving rotation, $M$. Our analysis of experimental data from Zhu et al. \cite{zhu_mechanosensitive_2024} support this prediction, but more direct measurements are needed. If this prediction is correct, it would have major implications for current models of bacterial chemotaxis. In particular, the steep motor CW bias response curves measured by Cluzel et al. \cite{cluzel_ultrasensitive_2000} and Yuan et al. \cite{yuan_ultrasensitivity_2013} used motors attached to 0.5-$\mathrm{\mu m}$ and 1-$\mathrm{\mu m}$ beads, respectively. These are relatively large loads, leading to higher stator numbers and potentially higher Hill coefficients than cells swimming in aqueous media \cite{niu_flagellar_2023}.

Although GMC can generate high cooperativity without conformational spread (nearest-neighbor interactions), which was essential in previous models, real motors may use a combination of both. FliG subunits in the C-ring are connected by a spring-shaped coil of FliM-FliN repeats \cite{chang_molecular_2020, johnson_structural_2024, singh_cryoem_2024}, and having neighboring FliG's in opposing conformations may induce energetically-unfavorable bending in this spring. High-resolution structures of the C-ring also suggest that steric interactions may favor alignment of neighboring FliG subunits \cite{johnson_structural_2024, singh_cryoem_2024}. Here, we showed that nearest-neighbor interactions and GMC are synergistic: the former increases the maximum cooperativity \cite{owen_size_2023}, while the latter enables faster responses to inputs.

Several features of the BFM rotation switch were not included here. We lumped the effects of CheYp binding into the free energy bias $\Delta F$, but binding has its own dynamics. Including effects of ion motive force would allow direct comparisons with more experiments. The output torque-speed curves of \textit{E. coli} motors rotating CW versus CCW have different shapes \cite{yuan_asymmetry_2010}, which could affect local torques on FliG subunits. 
More detailed modeling \cite{van_albada_switching_2009, mora_modeling_2009, meacci_dynamics_2009, meacci_dynamics_2011, tu_design_2018, cao_modeling_2022} could relax some of our assumptions, such as strong MotA-FliG interactions and C-ring rigidity. 

The BFM is also adaptive: it adapts the number of FliM binding sites for CheYp, depending on rotation direction \cite{delalez_signal-dependent_2010, yuan_adaptation_2012, lele_mechanism_2012, delalez_stoichiometry_2014, branch_adaptive_2014}, and it adapts the number of stators driving rotation, depending on the external load \cite{blair_restoration_1988, lele_dynamics_2013, tipping_load-dependent_2013, tusk_subunit_2018,  wadhwa_torque-dependent_2019, nirody_load-dependent_2019, wadhwa_mechanosensitive_2021, wadhwa_multi-state_2022}. 
GMC is a starting point for understanding how chemical and mechanical inputs are integrated into a functional switch response across a range of intracellular and environmental conditions \cite{antani_mechanosensitive_2021}.

Our results raise the possibility that dissipative mechanical forces may be leveraged for more-subtly dissipative responses to inputs in other biological contexts. A potential example is bidirectional cargo transport along microtubules \cite{gross_hither_2004, welte_bidirectional_2004,  hancock_bidirectional_2014}. Ongoing work \cite{muller_tug--war_2008, soppina_tug--war_2009, muller_bidirectional_2010, kunwar_mechanical_2011, hancock_bidirectional_2014, dsouza_vesicles_2023} suggests opposing motors attached to the same cargo undergo a tug of war that, together with force-dependent detachment rates from microtubules, could make transport highly sensitive to chemical regulation.

\begin{acknowledgments}
	{\it Acknowledgments ---} YT was partially supported by NIH grant R35GM137134. HHM thanks David Hathcock and Qiwei Yu for advice about computing transition paths and dissipation rates from simulations; Matthew Leighton for very helpful comments on the manuscript; and the Biophysical Modeling Group at the Flatiron Institute for feedback. HHM and YT thank Junhua Yuan and Rui He for the experimental measurements of CW bias in Fig. \ref{fig:data}. YT would like to thank the Center for Computational Biology at the Flatiron Institute for hospitality while a portion of this work was carried out.
\end{acknowledgments}

\bibliography{references}

\begin{thebibliography}{78}%
\makeatletter
\providecommand \@ifxundefined [1]{%
 \@ifx{#1\undefined}
}%
\providecommand \@ifnum [1]{%
 \ifnum #1\expandafter \@firstoftwo
 \else \expandafter \@secondoftwo
 \fi
}%
\providecommand \@ifx [1]{%
 \ifx #1\expandafter \@firstoftwo
 \else \expandafter \@secondoftwo
 \fi
}%
\providecommand \natexlab [1]{#1}%
\providecommand \enquote  [1]{``#1''}%
\providecommand \bibnamefont  [1]{#1}%
\providecommand \bibfnamefont [1]{#1}%
\providecommand \citenamefont [1]{#1}%
\providecommand \href@noop [0]{\@secondoftwo}%
\providecommand \href [0]{\begingroup \@sanitize@url \@href}%
\providecommand \@href[1]{\@@startlink{#1}\@@href}%
\providecommand \@@href[1]{\endgroup#1\@@endlink}%
\providecommand \@sanitize@url [0]{\catcode `\\12\catcode `\$12\catcode
  `\&12\catcode `\#12\catcode `\^12\catcode `\_12\catcode `\%12\relax}%
\providecommand \@@startlink[1]{}%
\providecommand \@@endlink[0]{}%
\providecommand \url  [0]{\begingroup\@sanitize@url \@url }%
\providecommand \@url [1]{\endgroup\@href {#1}{\urlprefix }}%
\providecommand \urlprefix  [0]{URL }%
\providecommand \Eprint [0]{\href }%
\providecommand \doibase [0]{https://doi.org/}%
\providecommand \selectlanguage [0]{\@gobble}%
\providecommand \bibinfo  [0]{\@secondoftwo}%
\providecommand \bibfield  [0]{\@secondoftwo}%
\providecommand \translation [1]{[#1]}%
\providecommand \BibitemOpen [0]{}%
\providecommand \bibitemStop [0]{}%
\providecommand \bibitemNoStop [0]{.\EOS\space}%
\providecommand \EOS [0]{\spacefactor3000\relax}%
\providecommand \BibitemShut  [1]{\csname bibitem#1\endcsname}%
\let\auto@bib@innerbib\@empty
\bibitem [{\citenamefont {Armitage}\ and\ \citenamefont
  {Berry}(2020)}]{armitage_assembly_2020}%
  \BibitemOpen
  \bibfield  {author} {\bibinfo {author} {\bibfnamefont {J.~P.}\ \bibnamefont
  {Armitage}}\ and\ \bibinfo {author} {\bibfnamefont {R.~M.}\ \bibnamefont
  {Berry}},\ }\bibfield  {title} {\bibinfo {title} {Assembly and {Dynamics} of
  the {Bacterial} {Flagellum}},\ }\href
  {https://doi.org/10.1146/annurev-micro-090816-093411} {\bibfield  {journal}
  {\bibinfo  {journal} {Annual Review of Microbiology}\ }\textbf {\bibinfo
  {volume} {74}},\ \bibinfo {pages} {181} (\bibinfo {year} {2020})}\BibitemShut
  {NoStop}%
\bibitem [{\citenamefont {Wadhwa}\ and\ \citenamefont
  {Berg}(2021)}]{wadhwa_bacterial_2021}%
  \BibitemOpen
  \bibfield  {author} {\bibinfo {author} {\bibfnamefont {N.}~\bibnamefont
  {Wadhwa}}\ and\ \bibinfo {author} {\bibfnamefont {H.~C.}\ \bibnamefont
  {Berg}},\ }\bibfield  {title} {\bibinfo {title} {Bacterial motility:
  machinery and mechanisms},\ }\href
  {https://doi.org/10.1038/s41579-021-00626-4} {\bibfield  {journal} {\bibinfo
  {journal} {Nature Reviews Microbiology}\ ,\ \bibinfo {pages} {1}} (\bibinfo
  {year} {2021})}\BibitemShut {NoStop}%
\bibitem [{\citenamefont {Guo}\ and\ \citenamefont
  {Liu}(2022)}]{guo_bacterial_2022}%
  \BibitemOpen
  \bibfield  {author} {\bibinfo {author} {\bibfnamefont {S.}~\bibnamefont
  {Guo}}\ and\ \bibinfo {author} {\bibfnamefont {J.}~\bibnamefont {Liu}},\
  }\bibfield  {title} {\bibinfo {title} {The {Bacterial} {Flagellar} {Motor}:
  {Insights} {Into} {Torque} {Generation}, {Rotational} {Switching}, and
  {Mechanosensing}},\ }\href
  {https://www.frontiersin.org/articles/10.3389/fmicb.2022.911114} {\bibfield
  {journal} {\bibinfo  {journal} {Frontiers in Microbiology}\ }\textbf
  {\bibinfo {volume} {13}} (\bibinfo {year} {2022})}\BibitemShut {NoStop}%
\bibitem [{\citenamefont {Hu}\ \emph {et~al.}(2022)\citenamefont {Hu},
  \citenamefont {Santiveri}, \citenamefont {Wadhwa}, \citenamefont {Berg},
  \citenamefont {Erhardt},\ and\ \citenamefont {Taylor}}]{hu_structural_2022}%
  \BibitemOpen
  \bibfield  {author} {\bibinfo {author} {\bibfnamefont {H.}~\bibnamefont
  {Hu}}, \bibinfo {author} {\bibfnamefont {M.}~\bibnamefont {Santiveri}},
  \bibinfo {author} {\bibfnamefont {N.}~\bibnamefont {Wadhwa}}, \bibinfo
  {author} {\bibfnamefont {H.~C.}\ \bibnamefont {Berg}}, \bibinfo {author}
  {\bibfnamefont {M.}~\bibnamefont {Erhardt}},\ and\ \bibinfo {author}
  {\bibfnamefont {N.~M.~I.}\ \bibnamefont {Taylor}},\ }\bibfield  {title}
  {\bibinfo {title} {Structural basis of torque generation in the
  bi-directional bacterial flagellar motor},\ }\href
  {https://doi.org/10.1016/j.tibs.2021.06.005} {\bibfield  {journal} {\bibinfo
  {journal} {Trends in Biochemical Sciences}\ }\bibinfo {series} {Special
  {Issue}: {Pushing} boundaries of cryo-{EM}},\ \textbf {\bibinfo {volume}
  {47}},\ \bibinfo {pages} {160} (\bibinfo {year} {2022})}\BibitemShut
  {NoStop}%
\bibitem [{\citenamefont {Berg}\ and\ \citenamefont
  {Brown}(1972)}]{berg_chemotaxis_1972}%
  \BibitemOpen
  \bibfield  {author} {\bibinfo {author} {\bibfnamefont {H.~C.}\ \bibnamefont
  {Berg}}\ and\ \bibinfo {author} {\bibfnamefont {D.~A.}\ \bibnamefont
  {Brown}},\ }\bibfield  {title} {\bibinfo {title} {Chemotaxis in {Escherichia}
  coli analysed by {Three}-dimensional {Tracking}},\ }\href
  {https://doi.org/10.1038/239500a0} {\bibfield  {journal} {\bibinfo  {journal}
  {Nature}\ }\textbf {\bibinfo {volume} {239}},\ \bibinfo {pages} {500}
  (\bibinfo {year} {1972})}\BibitemShut {NoStop}%
\bibitem [{\citenamefont {Grognot}\ and\ \citenamefont
  {Taute}(2021)}]{grognot_more_2021}%
  \BibitemOpen
  \bibfield  {author} {\bibinfo {author} {\bibfnamefont {M.}~\bibnamefont
  {Grognot}}\ and\ \bibinfo {author} {\bibfnamefont {K.~M.}\ \bibnamefont
  {Taute}},\ }\bibfield  {title} {\bibinfo {title} {More than propellers: how
  flagella shape bacterial motility behaviors},\ }\href
  {https://doi.org/10.1016/j.mib.2021.02.005} {\bibfield  {journal} {\bibinfo
  {journal} {Current Opinion in Microbiology}\ }\textbf {\bibinfo {volume}
  {61}},\ \bibinfo {pages} {73} (\bibinfo {year} {2021})}\BibitemShut {NoStop}%
\bibitem [{\citenamefont {Chen}\ \emph {et~al.}(2011)\citenamefont {Chen},
  \citenamefont {Beeby}, \citenamefont {Murphy}, \citenamefont {Leadbetter},
  \citenamefont {Hendrixson}, \citenamefont {Briegel}, \citenamefont {Li},
  \citenamefont {Shi}, \citenamefont {Tocheva}, \citenamefont {Müller},
  \citenamefont {Dobro},\ and\ \citenamefont {Jensen}}]{chen_structural_2011}%
  \BibitemOpen
  \bibfield  {author} {\bibinfo {author} {\bibfnamefont {S.}~\bibnamefont
  {Chen}}, \bibinfo {author} {\bibfnamefont {M.}~\bibnamefont {Beeby}},
  \bibinfo {author} {\bibfnamefont {G.~E.}\ \bibnamefont {Murphy}}, \bibinfo
  {author} {\bibfnamefont {J.~R.}\ \bibnamefont {Leadbetter}}, \bibinfo
  {author} {\bibfnamefont {D.~R.}\ \bibnamefont {Hendrixson}}, \bibinfo
  {author} {\bibfnamefont {A.}~\bibnamefont {Briegel}}, \bibinfo {author}
  {\bibfnamefont {Z.}~\bibnamefont {Li}}, \bibinfo {author} {\bibfnamefont
  {J.}~\bibnamefont {Shi}}, \bibinfo {author} {\bibfnamefont {E.~I.}\
  \bibnamefont {Tocheva}}, \bibinfo {author} {\bibfnamefont {A.}~\bibnamefont
  {Müller}}, \bibinfo {author} {\bibfnamefont {M.~J.}\ \bibnamefont {Dobro}},\
  and\ \bibinfo {author} {\bibfnamefont {G.~J.}\ \bibnamefont {Jensen}},\
  }\bibfield  {title} {\bibinfo {title} {Structural diversity of bacterial
  flagellar motors},\ }\href {https://doi.org/10.1038/emboj.2011.186}
  {\bibfield  {journal} {\bibinfo  {journal} {The EMBO Journal}\ }\textbf
  {\bibinfo {volume} {30}},\ \bibinfo {pages} {2972} (\bibinfo {year}
  {2011})}\BibitemShut {NoStop}%
\bibitem [{\citenamefont {Zhao}\ \emph {et~al.}(2014)\citenamefont {Zhao},
  \citenamefont {Norris},\ and\ \citenamefont {Liu}}]{zhao_molecular_2014}%
  \BibitemOpen
  \bibfield  {author} {\bibinfo {author} {\bibfnamefont {X.}~\bibnamefont
  {Zhao}}, \bibinfo {author} {\bibfnamefont {S.~J.}\ \bibnamefont {Norris}},\
  and\ \bibinfo {author} {\bibfnamefont {J.}~\bibnamefont {Liu}},\ }\bibfield
  {title} {\bibinfo {title} {Molecular {Architecture} of the {Bacterial}
  {Flagellar} {Motor} in {Cells}},\ }\href {https://doi.org/10.1021/bi500059y}
  {\bibfield  {journal} {\bibinfo  {journal} {Biochemistry}\ }\textbf {\bibinfo
  {volume} {53}},\ \bibinfo {pages} {4323} (\bibinfo {year}
  {2014})}\BibitemShut {NoStop}%
\bibitem [{\citenamefont {Carroll}\ and\ \citenamefont
  {Liu}(2020)}]{carroll_structural_2020}%
  \BibitemOpen
  \bibfield  {author} {\bibinfo {author} {\bibfnamefont {B.~L.}\ \bibnamefont
  {Carroll}}\ and\ \bibinfo {author} {\bibfnamefont {J.}~\bibnamefont {Liu}},\
  }\bibfield  {title} {\bibinfo {title} {Structural {Conservation} and
  {Adaptation} of the {Bacterial} {Flagella} {Motor}},\ }\href
  {https://doi.org/10.3390/biom10111492} {\bibfield  {journal} {\bibinfo
  {journal} {Biomolecules}\ }\textbf {\bibinfo {volume} {10}},\ \bibinfo
  {pages} {1492} (\bibinfo {year} {2020})}\BibitemShut {NoStop}%
\bibitem [{\citenamefont {Cluzel}\ \emph {et~al.}(2000)\citenamefont {Cluzel},
  \citenamefont {Surette},\ and\ \citenamefont
  {Leibler}}]{cluzel_ultrasensitive_2000}%
  \BibitemOpen
  \bibfield  {author} {\bibinfo {author} {\bibfnamefont {P.}~\bibnamefont
  {Cluzel}}, \bibinfo {author} {\bibfnamefont {M.}~\bibnamefont {Surette}},\
  and\ \bibinfo {author} {\bibfnamefont {S.}~\bibnamefont {Leibler}},\
  }\bibfield  {title} {\bibinfo {title} {An {Ultrasensitive} {Bacterial}
  {Motor} {Revealed} by {Monitoring} {Signaling} {Proteins} in {Single}
  {Cells}},\ }\href {https://doi.org/10.1126/science.287.5458.1652} {\bibfield
  {journal} {\bibinfo  {journal} {Science}\ }\textbf {\bibinfo {volume}
  {287}},\ \bibinfo {pages} {1652} (\bibinfo {year} {2000})}\BibitemShut
  {NoStop}%
\bibitem [{\citenamefont {Yuan}\ and\ \citenamefont
  {Berg}(2013)}]{yuan_ultrasensitivity_2013}%
  \BibitemOpen
  \bibfield  {author} {\bibinfo {author} {\bibfnamefont {J.}~\bibnamefont
  {Yuan}}\ and\ \bibinfo {author} {\bibfnamefont {H.~C.}\ \bibnamefont
  {Berg}},\ }\bibfield  {title} {\bibinfo {title} {Ultrasensitivity of an
  {Adaptive} {Bacterial} {Motor}},\ }\href
  {https://doi.org/10.1016/j.jmb.2013.02.016} {\bibfield  {journal} {\bibinfo
  {journal} {Journal of Molecular Biology}\ }\textbf {\bibinfo {volume}
  {425}},\ \bibinfo {pages} {1760} (\bibinfo {year} {2013})}\BibitemShut
  {NoStop}%
\bibitem [{\citenamefont {Thomas}\ \emph {et~al.}(1999)\citenamefont {Thomas},
  \citenamefont {Morgan},\ and\ \citenamefont
  {DeRosier}}]{thomas_rotational_1999}%
  \BibitemOpen
  \bibfield  {author} {\bibinfo {author} {\bibfnamefont {D.~R.}\ \bibnamefont
  {Thomas}}, \bibinfo {author} {\bibfnamefont {D.~G.}\ \bibnamefont {Morgan}},\
  and\ \bibinfo {author} {\bibfnamefont {D.~J.}\ \bibnamefont {DeRosier}},\
  }\bibfield  {title} {\bibinfo {title} {Rotational symmetry of the {C} ring
  and a mechanism for the flagellar rotary motor},\ }\href
  {https://doi.org/10.1073/pnas.96.18.10134} {\bibfield  {journal} {\bibinfo
  {journal} {Proceedings of the National Academy of Sciences}\ }\textbf
  {\bibinfo {volume} {96}},\ \bibinfo {pages} {10134} (\bibinfo {year}
  {1999})}\BibitemShut {NoStop}%
\bibitem [{\citenamefont {Duke}\ \emph {et~al.}(2001)\citenamefont {Duke},
  \citenamefont {Le~Novère},\ and\ \citenamefont
  {Bray}}]{duke_conformational_2001}%
  \BibitemOpen
  \bibfield  {author} {\bibinfo {author} {\bibfnamefont {T.~A.~J.}\
  \bibnamefont {Duke}}, \bibinfo {author} {\bibfnamefont {N.}~\bibnamefont
  {Le~Novère}},\ and\ \bibinfo {author} {\bibfnamefont {D.}~\bibnamefont
  {Bray}},\ }\bibfield  {title} {\bibinfo {title} {Conformational spread in a
  ring of proteins: a stochastic approach to allostery},\ }\href
  {https://doi.org/10.1006/jmbi.2001.4610} {\bibfield  {journal} {\bibinfo
  {journal} {Journal of Molecular Biology}\ }\textbf {\bibinfo {volume}
  {308}},\ \bibinfo {pages} {541} (\bibinfo {year} {2001})}\BibitemShut
  {NoStop}%
\bibitem [{\citenamefont {Bai}\ \emph {et~al.}(2010)\citenamefont {Bai},
  \citenamefont {Branch}, \citenamefont {Nicolau}, \citenamefont {Pilizota},
  \citenamefont {Steel}, \citenamefont {Maini},\ and\ \citenamefont
  {Berry}}]{bai_conformational_2010}%
  \BibitemOpen
  \bibfield  {author} {\bibinfo {author} {\bibfnamefont {F.}~\bibnamefont
  {Bai}}, \bibinfo {author} {\bibfnamefont {R.~W.}\ \bibnamefont {Branch}},
  \bibinfo {author} {\bibfnamefont {D.~V.}\ \bibnamefont {Nicolau}}, \bibinfo
  {author} {\bibfnamefont {T.}~\bibnamefont {Pilizota}}, \bibinfo {author}
  {\bibfnamefont {B.~C.}\ \bibnamefont {Steel}}, \bibinfo {author}
  {\bibfnamefont {P.~K.}\ \bibnamefont {Maini}},\ and\ \bibinfo {author}
  {\bibfnamefont {R.~M.}\ \bibnamefont {Berry}},\ }\bibfield  {title} {\bibinfo
  {title} {Conformational {Spread} as a {Mechanism} for {Cooperativity} in the
  {Bacterial} {Flagellar} {Switch}},\ }\href
  {https://doi.org/10.1126/science.1182105} {\bibfield  {journal} {\bibinfo
  {journal} {Science}\ }\textbf {\bibinfo {volume} {327}},\ \bibinfo {pages}
  {685} (\bibinfo {year} {2010})}\BibitemShut {NoStop}%
\bibitem [{\citenamefont {Korobkova}\ \emph {et~al.}(2006)\citenamefont
  {Korobkova}, \citenamefont {Emonet}, \citenamefont {Park},\ and\
  \citenamefont {Cluzel}}]{korobkova_hidden_2006}%
  \BibitemOpen
  \bibfield  {author} {\bibinfo {author} {\bibfnamefont {E.~A.}\ \bibnamefont
  {Korobkova}}, \bibinfo {author} {\bibfnamefont {T.}~\bibnamefont {Emonet}},
  \bibinfo {author} {\bibfnamefont {H.}~\bibnamefont {Park}},\ and\ \bibinfo
  {author} {\bibfnamefont {P.}~\bibnamefont {Cluzel}},\ }\bibfield  {title}
  {\bibinfo {title} {Hidden {Stochastic} {Nature} of a {Single} {Bacterial}
  {Motor}},\ }\href {https://doi.org/10.1103/PhysRevLett.96.058105} {\bibfield
  {journal} {\bibinfo  {journal} {Physical Review Letters}\ }\textbf {\bibinfo
  {volume} {96}},\ \bibinfo {pages} {058105} (\bibinfo {year}
  {2006})}\BibitemShut {NoStop}%
\bibitem [{\citenamefont {Wang}\ \emph {et~al.}(2017)\citenamefont {Wang},
  \citenamefont {Shi}, \citenamefont {He}, \citenamefont {Wang}, \citenamefont
  {Zhang},\ and\ \citenamefont {Yuan}}]{wang_non-equilibrium_2017}%
  \BibitemOpen
  \bibfield  {author} {\bibinfo {author} {\bibfnamefont {F.}~\bibnamefont
  {Wang}}, \bibinfo {author} {\bibfnamefont {H.}~\bibnamefont {Shi}}, \bibinfo
  {author} {\bibfnamefont {R.}~\bibnamefont {He}}, \bibinfo {author}
  {\bibfnamefont {R.}~\bibnamefont {Wang}}, \bibinfo {author} {\bibfnamefont
  {R.}~\bibnamefont {Zhang}},\ and\ \bibinfo {author} {\bibfnamefont
  {J.}~\bibnamefont {Yuan}},\ }\bibfield  {title} {\bibinfo {title}
  {Non-equilibrium effect in the allosteric regulation of the bacterial
  flagellar switch},\ }\href {https://doi.org/10.1038/nphys4081} {\bibfield
  {journal} {\bibinfo  {journal} {Nature Physics}\ }\textbf {\bibinfo {volume}
  {13}},\ \bibinfo {pages} {710} (\bibinfo {year} {2017})}\BibitemShut
  {NoStop}%
\bibitem [{\citenamefont {Tu}(2017)}]{tu_driven_2017}%
  \BibitemOpen
  \bibfield  {author} {\bibinfo {author} {\bibfnamefont {Y.}~\bibnamefont
  {Tu}},\ }\bibfield  {title} {\bibinfo {title} {Driven to peak},\ }\href
  {https://doi.org/10.1038/nphys4094} {\bibfield  {journal} {\bibinfo
  {journal} {Nature Physics}\ }\textbf {\bibinfo {volume} {13}},\ \bibinfo
  {pages} {631} (\bibinfo {year} {2017})}\BibitemShut {NoStop}%
\bibitem [{\citenamefont {Tu}(2008)}]{tu_nonequilibrium_2008}%
  \BibitemOpen
  \bibfield  {author} {\bibinfo {author} {\bibfnamefont {Y.}~\bibnamefont
  {Tu}},\ }\bibfield  {title} {\bibinfo {title} {The nonequilibrium mechanism
  for ultrasensitivity in a biological switch: {Sensing} by {Maxwell}'s
  demons},\ }\href {https://doi.org/10.1073/pnas.0804641105} {\bibfield
  {journal} {\bibinfo  {journal} {Proceedings of the National Academy of
  Sciences}\ }\textbf {\bibinfo {volume} {105}},\ \bibinfo {pages} {11737}
  (\bibinfo {year} {2008})}\BibitemShut {NoStop}%
\bibitem [{\citenamefont {Yuan}\ \emph {et~al.}(2009)\citenamefont {Yuan},
  \citenamefont {Fahrner},\ and\ \citenamefont {Berg}}]{yuan_switching_2009}%
  \BibitemOpen
  \bibfield  {author} {\bibinfo {author} {\bibfnamefont {J.}~\bibnamefont
  {Yuan}}, \bibinfo {author} {\bibfnamefont {K.~A.}\ \bibnamefont {Fahrner}},\
  and\ \bibinfo {author} {\bibfnamefont {H.~C.}\ \bibnamefont {Berg}},\
  }\bibfield  {title} {\bibinfo {title} {Switching of the {Bacterial}
  {Flagellar} {Motor} {Near} {Zero} {Load}},\ }\href
  {https://doi.org/10.1016/j.jmb.2009.05.039} {\bibfield  {journal} {\bibinfo
  {journal} {Journal of Molecular Biology}\ }\textbf {\bibinfo {volume}
  {390}},\ \bibinfo {pages} {394} (\bibinfo {year} {2009})}\BibitemShut
  {NoStop}%
\bibitem [{\citenamefont {Bai}\ \emph {et~al.}(2012)\citenamefont {Bai},
  \citenamefont {Minamino}, \citenamefont {Wu}, \citenamefont {Namba},\ and\
  \citenamefont {Xing}}]{bai_coupling_2012}%
  \BibitemOpen
  \bibfield  {author} {\bibinfo {author} {\bibfnamefont {F.}~\bibnamefont
  {Bai}}, \bibinfo {author} {\bibfnamefont {T.}~\bibnamefont {Minamino}},
  \bibinfo {author} {\bibfnamefont {Z.}~\bibnamefont {Wu}}, \bibinfo {author}
  {\bibfnamefont {K.}~\bibnamefont {Namba}},\ and\ \bibinfo {author}
  {\bibfnamefont {J.}~\bibnamefont {Xing}},\ }\bibfield  {title} {\bibinfo
  {title} {Coupling between {Switching} {Regulation} and {Torque} {Generation}
  in {Bacterial} {Flagellar} {Motor}},\ }\href
  {https://doi.org/10.1103/PhysRevLett.108.178105} {\bibfield  {journal}
  {\bibinfo  {journal} {Physical Review Letters}\ }\textbf {\bibinfo {volume}
  {108}},\ \bibinfo {pages} {178105} (\bibinfo {year} {2012})}\BibitemShut
  {NoStop}%
\bibitem [{\citenamefont {Wang}\ \emph {et~al.}(2021)\citenamefont {Wang},
  \citenamefont {Niu}, \citenamefont {Zhang},\ and\ \citenamefont
  {Yuan}}]{wang_dynamics_2021}%
  \BibitemOpen
  \bibfield  {author} {\bibinfo {author} {\bibfnamefont {B.}~\bibnamefont
  {Wang}}, \bibinfo {author} {\bibfnamefont {Y.}~\bibnamefont {Niu}}, \bibinfo
  {author} {\bibfnamefont {R.}~\bibnamefont {Zhang}},\ and\ \bibinfo {author}
  {\bibfnamefont {J.}~\bibnamefont {Yuan}},\ }\bibfield  {title} {\bibinfo
  {title} {Dynamics of {Switching} at {Stall} {Reveals} {Nonequilibrium}
  {Mechanism} in the {Allosteric} {Regulation} of the {Bacterial} {Flagellar}
  {Switch}},\ }\href {https://doi.org/10.1103/PhysRevLett.127.268101}
  {\bibfield  {journal} {\bibinfo  {journal} {Physical Review Letters}\
  }\textbf {\bibinfo {volume} {127}},\ \bibinfo {pages} {268101} (\bibinfo
  {year} {2021})}\BibitemShut {NoStop}%
\bibitem [{\citenamefont {Zhu}\ \emph {et~al.}(2024)\citenamefont {Zhu},
  \citenamefont {He}, \citenamefont {Zhang},\ and\ \citenamefont
  {Yuan}}]{zhu_mechanosensitive_2024}%
  \BibitemOpen
  \bibfield  {author} {\bibinfo {author} {\bibfnamefont {S.}~\bibnamefont
  {Zhu}}, \bibinfo {author} {\bibfnamefont {R.}~\bibnamefont {He}}, \bibinfo
  {author} {\bibfnamefont {R.}~\bibnamefont {Zhang}},\ and\ \bibinfo {author}
  {\bibfnamefont {J.}~\bibnamefont {Yuan}},\ }\bibfield  {title} {\bibinfo
  {title} {Mechanosensitive dose response of the bacterial flagellar motor},\
  }\href {https://doi.org/10.1103/PhysRevE.110.054402} {\bibfield  {journal}
  {\bibinfo  {journal} {Physical Review E}\ }\textbf {\bibinfo {volume}
  {110}},\ \bibinfo {pages} {054402} (\bibinfo {year} {2024})}\BibitemShut
  {NoStop}%
\bibitem [{\citenamefont {Deme}\ \emph {et~al.}(2020)\citenamefont {Deme},
  \citenamefont {Johnson}, \citenamefont {Vickery}, \citenamefont {Aron},
  \citenamefont {Monkhouse}, \citenamefont {Griffiths}, \citenamefont {James},
  \citenamefont {Berks}, \citenamefont {Coulton}, \citenamefont {Stansfeld},\
  and\ \citenamefont {Lea}}]{deme_structures_2020}%
  \BibitemOpen
  \bibfield  {author} {\bibinfo {author} {\bibfnamefont {J.~C.}\ \bibnamefont
  {Deme}}, \bibinfo {author} {\bibfnamefont {S.}~\bibnamefont {Johnson}},
  \bibinfo {author} {\bibfnamefont {O.}~\bibnamefont {Vickery}}, \bibinfo
  {author} {\bibfnamefont {A.}~\bibnamefont {Aron}}, \bibinfo {author}
  {\bibfnamefont {H.}~\bibnamefont {Monkhouse}}, \bibinfo {author}
  {\bibfnamefont {T.}~\bibnamefont {Griffiths}}, \bibinfo {author}
  {\bibfnamefont {R.~H.}\ \bibnamefont {James}}, \bibinfo {author}
  {\bibfnamefont {B.~C.}\ \bibnamefont {Berks}}, \bibinfo {author}
  {\bibfnamefont {J.~W.}\ \bibnamefont {Coulton}}, \bibinfo {author}
  {\bibfnamefont {P.~J.}\ \bibnamefont {Stansfeld}},\ and\ \bibinfo {author}
  {\bibfnamefont {S.~M.}\ \bibnamefont {Lea}},\ }\bibfield  {title} {\bibinfo
  {title} {Structures of the stator complex that drives rotation of the
  bacterial flagellum},\ }\href {https://doi.org/10.1038/s41564-020-0788-8}
  {\bibfield  {journal} {\bibinfo  {journal} {Nature Microbiology}\ }\textbf
  {\bibinfo {volume} {5}},\ \bibinfo {pages} {1553} (\bibinfo {year}
  {2020})}\BibitemShut {NoStop}%
\bibitem [{\citenamefont {Santiveri}\ \emph {et~al.}(2020)\citenamefont
  {Santiveri}, \citenamefont {Roa-Eguiara}, \citenamefont {Kühne},
  \citenamefont {Wadhwa}, \citenamefont {Hu}, \citenamefont {Berg},
  \citenamefont {Erhardt},\ and\ \citenamefont
  {Taylor}}]{santiveri_structure_2020}%
  \BibitemOpen
  \bibfield  {author} {\bibinfo {author} {\bibfnamefont {M.}~\bibnamefont
  {Santiveri}}, \bibinfo {author} {\bibfnamefont {A.}~\bibnamefont
  {Roa-Eguiara}}, \bibinfo {author} {\bibfnamefont {C.}~\bibnamefont {Kühne}},
  \bibinfo {author} {\bibfnamefont {N.}~\bibnamefont {Wadhwa}}, \bibinfo
  {author} {\bibfnamefont {H.}~\bibnamefont {Hu}}, \bibinfo {author}
  {\bibfnamefont {H.~C.}\ \bibnamefont {Berg}}, \bibinfo {author}
  {\bibfnamefont {M.}~\bibnamefont {Erhardt}},\ and\ \bibinfo {author}
  {\bibfnamefont {N.~M.~I.}\ \bibnamefont {Taylor}},\ }\bibfield  {title}
  {\bibinfo {title} {Structure and {Function} of {Stator} {Units} of the
  {Bacterial} {Flagellar} {Motor}},\ }\href
  {https://doi.org/10.1016/j.cell.2020.08.016} {\bibfield  {journal} {\bibinfo
  {journal} {Cell}\ }\textbf {\bibinfo {volume} {183}},\ \bibinfo {pages} {244}
  (\bibinfo {year} {2020})}\BibitemShut {NoStop}%
\bibitem [{\citenamefont {Johnson}\ \emph {et~al.}(2024)\citenamefont
  {Johnson}, \citenamefont {Deme}, \citenamefont {Furlong}, \citenamefont
  {Caesar}, \citenamefont {Chevance}, \citenamefont {Hughes},\ and\
  \citenamefont {Lea}}]{johnson_structural_2024}%
  \BibitemOpen
  \bibfield  {author} {\bibinfo {author} {\bibfnamefont {S.}~\bibnamefont
  {Johnson}}, \bibinfo {author} {\bibfnamefont {J.~C.}\ \bibnamefont {Deme}},
  \bibinfo {author} {\bibfnamefont {E.~J.}\ \bibnamefont {Furlong}}, \bibinfo
  {author} {\bibfnamefont {J.~J.~E.}\ \bibnamefont {Caesar}}, \bibinfo {author}
  {\bibfnamefont {F.~F.~V.}\ \bibnamefont {Chevance}}, \bibinfo {author}
  {\bibfnamefont {K.~T.}\ \bibnamefont {Hughes}},\ and\ \bibinfo {author}
  {\bibfnamefont {S.~M.}\ \bibnamefont {Lea}},\ }\bibfield  {title} {\bibinfo
  {title} {Structural basis of directional switching by the bacterial
  flagellum},\ }\href {https://doi.org/10.1038/s41564-024-01630-z} {\bibfield
  {journal} {\bibinfo  {journal} {Nature Microbiology}\ }\textbf {\bibinfo
  {volume} {9}},\ \bibinfo {pages} {1282} (\bibinfo {year} {2024})}\BibitemShut
  {NoStop}%
\bibitem [{\citenamefont {Chang}\ \emph {et~al.}(2020)\citenamefont {Chang},
  \citenamefont {Zhang}, \citenamefont {Carroll}, \citenamefont {Zhao},
  \citenamefont {Charon}, \citenamefont {Norris}, \citenamefont {Motaleb},
  \citenamefont {Li},\ and\ \citenamefont {Liu}}]{chang_molecular_2020}%
  \BibitemOpen
  \bibfield  {author} {\bibinfo {author} {\bibfnamefont {Y.}~\bibnamefont
  {Chang}}, \bibinfo {author} {\bibfnamefont {K.}~\bibnamefont {Zhang}},
  \bibinfo {author} {\bibfnamefont {B.~L.}\ \bibnamefont {Carroll}}, \bibinfo
  {author} {\bibfnamefont {X.}~\bibnamefont {Zhao}}, \bibinfo {author}
  {\bibfnamefont {N.~W.}\ \bibnamefont {Charon}}, \bibinfo {author}
  {\bibfnamefont {S.~J.}\ \bibnamefont {Norris}}, \bibinfo {author}
  {\bibfnamefont {M.~A.}\ \bibnamefont {Motaleb}}, \bibinfo {author}
  {\bibfnamefont {C.}~\bibnamefont {Li}},\ and\ \bibinfo {author}
  {\bibfnamefont {J.}~\bibnamefont {Liu}},\ }\bibfield  {title} {\bibinfo
  {title} {Molecular mechanism for rotational switching of the bacterial
  flagellar motor},\ }\href {https://doi.org/10.1038/s41594-020-0497-2}
  {\bibfield  {journal} {\bibinfo  {journal} {Nature Structural \& Molecular
  Biology}\ }\textbf {\bibinfo {volume} {27}},\ \bibinfo {pages} {1041}
  (\bibinfo {year} {2020})}\BibitemShut {NoStop}%
\bibitem [{\citenamefont {Carroll}\ \emph {et~al.}(2020)\citenamefont
  {Carroll}, \citenamefont {Nishikino}, \citenamefont {Guo}, \citenamefont
  {Zhu}, \citenamefont {Kojima}, \citenamefont {Homma},\ and\ \citenamefont
  {Liu}}]{carroll_flagellar_2020}%
  \BibitemOpen
  \bibfield  {author} {\bibinfo {author} {\bibfnamefont {B.~L.}\ \bibnamefont
  {Carroll}}, \bibinfo {author} {\bibfnamefont {T.}~\bibnamefont {Nishikino}},
  \bibinfo {author} {\bibfnamefont {W.}~\bibnamefont {Guo}}, \bibinfo {author}
  {\bibfnamefont {S.}~\bibnamefont {Zhu}}, \bibinfo {author} {\bibfnamefont
  {S.}~\bibnamefont {Kojima}}, \bibinfo {author} {\bibfnamefont
  {M.}~\bibnamefont {Homma}},\ and\ \bibinfo {author} {\bibfnamefont
  {J.}~\bibnamefont {Liu}},\ }\bibfield  {title} {\bibinfo {title} {The
  flagellar motor of {Vibrio} alginolyticus undergoes major structural
  remodeling during rotational switching},\ }\href
  {https://doi.org/10.7554/eLife.61446} {\bibfield  {journal} {\bibinfo
  {journal} {eLife}\ }\textbf {\bibinfo {volume} {9}},\ \bibinfo {pages}
  {e61446} (\bibinfo {year} {2020})}\BibitemShut {NoStop}%
\bibitem [{\citenamefont {Blair}\ and\ \citenamefont
  {Berg}(1988)}]{blair_restoration_1988}%
  \BibitemOpen
  \bibfield  {author} {\bibinfo {author} {\bibfnamefont {D.~F.}\ \bibnamefont
  {Blair}}\ and\ \bibinfo {author} {\bibfnamefont {H.~C.}\ \bibnamefont
  {Berg}},\ }\bibfield  {title} {\bibinfo {title} {Restoration of {Torque} in
  {Defective} {Flagellar} {Motors}},\ }\href
  {https://doi.org/10.1126/science.2849208} {\bibfield  {journal} {\bibinfo
  {journal} {Science}\ }\textbf {\bibinfo {volume} {242}},\ \bibinfo {pages}
  {1678} (\bibinfo {year} {1988})}\BibitemShut {NoStop}%
\bibitem [{\citenamefont {Lele}\ \emph {et~al.}(2013)\citenamefont {Lele},
  \citenamefont {Hosu},\ and\ \citenamefont {Berg}}]{lele_dynamics_2013}%
  \BibitemOpen
  \bibfield  {author} {\bibinfo {author} {\bibfnamefont {P.~P.}\ \bibnamefont
  {Lele}}, \bibinfo {author} {\bibfnamefont {B.~G.}\ \bibnamefont {Hosu}},\
  and\ \bibinfo {author} {\bibfnamefont {H.~C.}\ \bibnamefont {Berg}},\
  }\bibfield  {title} {\bibinfo {title} {Dynamics of mechanosensing in the
  bacterial flagellar motor},\ }\href {https://doi.org/10.1073/pnas.1305885110}
  {\bibfield  {journal} {\bibinfo  {journal} {Proceedings of the National
  Academy of Sciences}\ }\textbf {\bibinfo {volume} {110}},\ \bibinfo {pages}
  {11839} (\bibinfo {year} {2013})}\BibitemShut {NoStop}%
\bibitem [{\citenamefont {Tipping}\ \emph {et~al.}(2013)\citenamefont
  {Tipping}, \citenamefont {Delalez}, \citenamefont {Lim}, \citenamefont
  {Berry},\ and\ \citenamefont {Armitage}}]{tipping_load-dependent_2013}%
  \BibitemOpen
  \bibfield  {author} {\bibinfo {author} {\bibfnamefont {M.~J.}\ \bibnamefont
  {Tipping}}, \bibinfo {author} {\bibfnamefont {N.~J.}\ \bibnamefont
  {Delalez}}, \bibinfo {author} {\bibfnamefont {R.}~\bibnamefont {Lim}},
  \bibinfo {author} {\bibfnamefont {R.~M.}\ \bibnamefont {Berry}},\ and\
  \bibinfo {author} {\bibfnamefont {J.~P.}\ \bibnamefont {Armitage}},\
  }\bibfield  {title} {\bibinfo {title} {Load-{Dependent} {Assembly} of the
  {Bacterial} {Flagellar} {Motor}},\ }\href
  {https://doi.org/10.1128/mbio.00551-13} {\bibfield  {journal} {\bibinfo
  {journal} {mBio}\ }\textbf {\bibinfo {volume} {4}},\ \bibinfo {pages}
  {10.1128/mbio.00551} (\bibinfo {year} {2013})}\BibitemShut {NoStop}%
\bibitem [{\citenamefont {Tusk}\ \emph {et~al.}(2018)\citenamefont {Tusk},
  \citenamefont {Delalez},\ and\ \citenamefont {Berry}}]{tusk_subunit_2018}%
  \BibitemOpen
  \bibfield  {author} {\bibinfo {author} {\bibfnamefont {S.~E.}\ \bibnamefont
  {Tusk}}, \bibinfo {author} {\bibfnamefont {N.~J.}\ \bibnamefont {Delalez}},\
  and\ \bibinfo {author} {\bibfnamefont {R.~M.}\ \bibnamefont {Berry}},\
  }\bibfield  {title} {\bibinfo {title} {Subunit {Exchange} in {Protein}
  {Complexes}},\ }\href {https://doi.org/10.1016/j.jmb.2018.06.039} {\bibfield
  {journal} {\bibinfo  {journal} {Journal of Molecular Biology}\ }\bibinfo
  {series} {Plasticity of {Multi}-{Protein} {Complexes}},\ \textbf {\bibinfo
  {volume} {430}},\ \bibinfo {pages} {4557} (\bibinfo {year}
  {2018})}\BibitemShut {NoStop}%
\bibitem [{\citenamefont {Wadhwa}\ \emph {et~al.}(2019)\citenamefont {Wadhwa},
  \citenamefont {Phillips},\ and\ \citenamefont
  {Berg}}]{wadhwa_torque-dependent_2019}%
  \BibitemOpen
  \bibfield  {author} {\bibinfo {author} {\bibfnamefont {N.}~\bibnamefont
  {Wadhwa}}, \bibinfo {author} {\bibfnamefont {R.}~\bibnamefont {Phillips}},\
  and\ \bibinfo {author} {\bibfnamefont {H.~C.}\ \bibnamefont {Berg}},\
  }\bibfield  {title} {\bibinfo {title} {Torque-dependent remodeling of the
  bacterial flagellar motor},\ }\href {https://doi.org/10.1073/pnas.1904577116}
  {\bibfield  {journal} {\bibinfo  {journal} {Proceedings of the National
  Academy of Sciences}\ }\textbf {\bibinfo {volume} {116}},\ \bibinfo {pages}
  {11764} (\bibinfo {year} {2019})}\BibitemShut {NoStop}%
\bibitem [{\citenamefont {Nirody}\ \emph {et~al.}(2019)\citenamefont {Nirody},
  \citenamefont {Nord},\ and\ \citenamefont
  {Berry}}]{nirody_load-dependent_2019}%
  \BibitemOpen
  \bibfield  {author} {\bibinfo {author} {\bibfnamefont {J.~A.}\ \bibnamefont
  {Nirody}}, \bibinfo {author} {\bibfnamefont {A.~L.}\ \bibnamefont {Nord}},\
  and\ \bibinfo {author} {\bibfnamefont {R.~M.}\ \bibnamefont {Berry}},\
  }\bibfield  {title} {\bibinfo {title} {Load-dependent adaptation near zero
  load in the bacterial flagellar motor},\ }\href
  {https://doi.org/10.1098/rsif.2019.0300} {\bibfield  {journal} {\bibinfo
  {journal} {Journal of The Royal Society Interface}\ }\textbf {\bibinfo
  {volume} {16}},\ \bibinfo {pages} {20190300} (\bibinfo {year}
  {2019})}\BibitemShut {NoStop}%
\bibitem [{\citenamefont {Wadhwa}\ \emph {et~al.}(2021)\citenamefont {Wadhwa},
  \citenamefont {Tu},\ and\ \citenamefont
  {Berg}}]{wadhwa_mechanosensitive_2021}%
  \BibitemOpen
  \bibfield  {author} {\bibinfo {author} {\bibfnamefont {N.}~\bibnamefont
  {Wadhwa}}, \bibinfo {author} {\bibfnamefont {Y.}~\bibnamefont {Tu}},\ and\
  \bibinfo {author} {\bibfnamefont {H.~C.}\ \bibnamefont {Berg}},\ }\bibfield
  {title} {\bibinfo {title} {Mechanosensitive remodeling of the bacterial
  flagellar motor is independent of direction of rotation},\ }\href
  {https://doi.org/10.1073/pnas.2024608118} {\bibfield  {journal} {\bibinfo
  {journal} {Proceedings of the National Academy of Sciences}\ }\textbf
  {\bibinfo {volume} {118}},\ \bibinfo {pages} {e2024608118} (\bibinfo {year}
  {2021})}\BibitemShut {NoStop}%
\bibitem [{\citenamefont {Wadhwa}\ \emph {et~al.}(2022)\citenamefont {Wadhwa},
  \citenamefont {Sassi}, \citenamefont {Berg},\ and\ \citenamefont
  {Tu}}]{wadhwa_multi-state_2022}%
  \BibitemOpen
  \bibfield  {author} {\bibinfo {author} {\bibfnamefont {N.}~\bibnamefont
  {Wadhwa}}, \bibinfo {author} {\bibfnamefont {A.}~\bibnamefont {Sassi}},
  \bibinfo {author} {\bibfnamefont {H.~C.}\ \bibnamefont {Berg}},\ and\
  \bibinfo {author} {\bibfnamefont {Y.}~\bibnamefont {Tu}},\ }\bibfield
  {title} {\bibinfo {title} {A multi-state dynamic process confers
  mechano-adaptation to a biological nanomachine},\ }\href
  {https://doi.org/10.1038/s41467-022-33075-5} {\bibfield  {journal} {\bibinfo
  {journal} {Nature Communications}\ }\textbf {\bibinfo {volume} {13}},\
  \bibinfo {pages} {5327} (\bibinfo {year} {2022})}\BibitemShut {NoStop}%
\bibitem [{\citenamefont {Reid}\ \emph {et~al.}(2006)\citenamefont {Reid},
  \citenamefont {Leake}, \citenamefont {Chandler}, \citenamefont {Lo},
  \citenamefont {Armitage},\ and\ \citenamefont {Berry}}]{reid_maximum_2006}%
  \BibitemOpen
  \bibfield  {author} {\bibinfo {author} {\bibfnamefont {S.~W.}\ \bibnamefont
  {Reid}}, \bibinfo {author} {\bibfnamefont {M.~C.}\ \bibnamefont {Leake}},
  \bibinfo {author} {\bibfnamefont {J.~H.}\ \bibnamefont {Chandler}}, \bibinfo
  {author} {\bibfnamefont {C.-J.}\ \bibnamefont {Lo}}, \bibinfo {author}
  {\bibfnamefont {J.~P.}\ \bibnamefont {Armitage}},\ and\ \bibinfo {author}
  {\bibfnamefont {R.~M.}\ \bibnamefont {Berry}},\ }\bibfield  {title} {\bibinfo
  {title} {The maximum number of torque-generating units in the flagellar motor
  of {Escherichia} coli is at least 11},\ }\href
  {https://doi.org/10.1073/pnas.0509932103} {\bibfield  {journal} {\bibinfo
  {journal} {Proceedings of the National Academy of Sciences}\ }\textbf
  {\bibinfo {volume} {103}},\ \bibinfo {pages} {8066} (\bibinfo {year}
  {2006})}\BibitemShut {NoStop}%
\bibitem [{\citenamefont {Chang}\ \emph {et~al.}(2021)\citenamefont {Chang},
  \citenamefont {Carroll},\ and\ \citenamefont {Liu}}]{chang_structural_2021}%
  \BibitemOpen
  \bibfield  {author} {\bibinfo {author} {\bibfnamefont {Y.}~\bibnamefont
  {Chang}}, \bibinfo {author} {\bibfnamefont {B.~L.}\ \bibnamefont {Carroll}},\
  and\ \bibinfo {author} {\bibfnamefont {J.}~\bibnamefont {Liu}},\ }\bibfield
  {title} {\bibinfo {title} {Structural basis of bacterial flagellar motor
  rotation and switching},\ }\href {https://doi.org/10.1016/j.tim.2021.03.009}
  {\bibfield  {journal} {\bibinfo  {journal} {Trends in Microbiology}\ }\textbf
  {\bibinfo {volume} {29}},\ \bibinfo {pages} {1024} (\bibinfo {year}
  {2021})}\BibitemShut {NoStop}%
\bibitem [{\citenamefont {Bell}(1978)}]{bell_models_1978}%
  \BibitemOpen
  \bibfield  {author} {\bibinfo {author} {\bibfnamefont {G.~I.}\ \bibnamefont
  {Bell}},\ }\bibfield  {title} {\bibinfo {title} {Models for the {Specific}
  {Adhesion} of {Cells} to {Cells}},\ }\href
  {https://doi.org/10.1126/science.347575} {\bibfield  {journal} {\bibinfo
  {journal} {Science}\ }\textbf {\bibinfo {volume} {200}},\ \bibinfo {pages}
  {618} (\bibinfo {year} {1978})}\BibitemShut {NoStop}%
\bibitem [{\citenamefont {Wiita}\ \emph {et~al.}(2006)\citenamefont {Wiita},
  \citenamefont {Ainavarapu}, \citenamefont {Huang},\ and\ \citenamefont
  {Fernandez}}]{wiita_force-dependent_2006}%
  \BibitemOpen
  \bibfield  {author} {\bibinfo {author} {\bibfnamefont {A.~P.}\ \bibnamefont
  {Wiita}}, \bibinfo {author} {\bibfnamefont {S.~R.~K.}\ \bibnamefont
  {Ainavarapu}}, \bibinfo {author} {\bibfnamefont {H.~H.}\ \bibnamefont
  {Huang}},\ and\ \bibinfo {author} {\bibfnamefont {J.~M.}\ \bibnamefont
  {Fernandez}},\ }\bibfield  {title} {\bibinfo {title} {Force-dependent
  chemical kinetics of disulfide bond reduction observed with single-molecule
  techniques},\ }\href {https://doi.org/10.1073/pnas.0511035103} {\bibfield
  {journal} {\bibinfo  {journal} {Proceedings of the National Academy of
  Sciences}\ }\textbf {\bibinfo {volume} {103}},\ \bibinfo {pages} {7222}
  (\bibinfo {year} {2006})}\BibitemShut {NoStop}%
\bibitem [{\citenamefont {Ryu}\ \emph {et~al.}(2000)\citenamefont {Ryu},
  \citenamefont {Berry},\ and\ \citenamefont
  {Berg}}]{ryu_torque-generating_2000}%
  \BibitemOpen
  \bibfield  {author} {\bibinfo {author} {\bibfnamefont {W.~S.}\ \bibnamefont
  {Ryu}}, \bibinfo {author} {\bibfnamefont {R.~M.}\ \bibnamefont {Berry}},\
  and\ \bibinfo {author} {\bibfnamefont {H.~C.}\ \bibnamefont {Berg}},\
  }\bibfield  {title} {\bibinfo {title} {Torque-generating units of the
  flagellar motor of {Escherichia} coli have a high duty ratio},\ }\href
  {https://doi.org/10.1038/35000233} {\bibfield  {journal} {\bibinfo  {journal}
  {Nature}\ }\textbf {\bibinfo {volume} {403}},\ \bibinfo {pages} {444}
  (\bibinfo {year} {2000})}\BibitemShut {NoStop}%
\bibitem [{\citenamefont {Yuan}\ and\ \citenamefont
  {Berg}(2008)}]{yuan_resurrection_2008}%
  \BibitemOpen
  \bibfield  {author} {\bibinfo {author} {\bibfnamefont {J.}~\bibnamefont
  {Yuan}}\ and\ \bibinfo {author} {\bibfnamefont {H.~C.}\ \bibnamefont
  {Berg}},\ }\bibfield  {title} {\bibinfo {title} {Resurrection of the
  flagellar rotary motor near zero load},\ }\href
  {https://doi.org/10.1073/pnas.0711539105} {\bibfield  {journal} {\bibinfo
  {journal} {Proceedings of the National Academy of Sciences}\ }\textbf
  {\bibinfo {volume} {105}},\ \bibinfo {pages} {1182} (\bibinfo {year}
  {2008})}\BibitemShut {NoStop}%
\bibitem [{\citenamefont {Nakamura}\ \emph {et~al.}(2009)\citenamefont
  {Nakamura}, \citenamefont {Kami-ike}, \citenamefont {Yokota}, \citenamefont
  {Kudo}, \citenamefont {Minamino},\ and\ \citenamefont
  {Namba}}]{nakamura_effect_2009}%
  \BibitemOpen
  \bibfield  {author} {\bibinfo {author} {\bibfnamefont {S.}~\bibnamefont
  {Nakamura}}, \bibinfo {author} {\bibfnamefont {N.}~\bibnamefont {Kami-ike}},
  \bibinfo {author} {\bibfnamefont {J.-i.~P.}\ \bibnamefont {Yokota}}, \bibinfo
  {author} {\bibfnamefont {S.}~\bibnamefont {Kudo}}, \bibinfo {author}
  {\bibfnamefont {T.}~\bibnamefont {Minamino}},\ and\ \bibinfo {author}
  {\bibfnamefont {K.}~\bibnamefont {Namba}},\ }\bibfield  {title} {\bibinfo
  {title} {Effect of {Intracellular} {pH} on the {Torque}–{Speed}
  {Relationship} of {Bacterial} {Proton}-{Driven} {Flagellar} {Motor}},\ }\href
  {https://doi.org/10.1016/j.jmb.2008.12.034} {\bibfield  {journal} {\bibinfo
  {journal} {Journal of Molecular Biology}\ }\textbf {\bibinfo {volume}
  {386}},\ \bibinfo {pages} {332} (\bibinfo {year} {2009})}\BibitemShut
  {NoStop}%
\bibitem [{\citenamefont {Niu}\ \emph {et~al.}(2023)\citenamefont {Niu},
  \citenamefont {Zhang},\ and\ \citenamefont {Yuan}}]{niu_flagellar_2023}%
  \BibitemOpen
  \bibfield  {author} {\bibinfo {author} {\bibfnamefont {Y.}~\bibnamefont
  {Niu}}, \bibinfo {author} {\bibfnamefont {R.}~\bibnamefont {Zhang}},\ and\
  \bibinfo {author} {\bibfnamefont {J.}~\bibnamefont {Yuan}},\ }\bibfield
  {title} {\bibinfo {title} {Flagellar motors of swimming bacteria contain an
  incomplete set of stator units to ensure robust motility},\ }\href
  {https://doi.org/10.1126/sciadv.adi6724} {\bibfield  {journal} {\bibinfo
  {journal} {Science Advances}\ }\textbf {\bibinfo {volume} {9}},\ \bibinfo
  {pages} {eadi6724} (\bibinfo {year} {2023})}\BibitemShut {NoStop}%
\bibitem [{\citenamefont {Gardiner}(2009)}]{gardiner_stochastic_2009}%
  \BibitemOpen
  \bibfield  {author} {\bibinfo {author} {\bibfnamefont {C.}~\bibnamefont
  {Gardiner}},\ }\href {https://link.springer.com/book/9783540707127} {\emph
  {\bibinfo {title} {Stochastic {Methods}: {A} {Handbook} for the {Natural} and
  {Social} {Sciences}}}},\ \bibinfo {edition} {4th}\ ed.\ (\bibinfo
  {publisher} {Springer},\ \bibinfo {address} {Berlin, Heidelberg},\ \bibinfo
  {year} {2009})\BibitemShut {NoStop}%
\bibitem [{\citenamefont {Monod}\ \emph {et~al.}(1965)\citenamefont {Monod},
  \citenamefont {Wyman},\ and\ \citenamefont {Changeux}}]{monod_nature_1965}%
  \BibitemOpen
  \bibfield  {author} {\bibinfo {author} {\bibfnamefont {J.}~\bibnamefont
  {Monod}}, \bibinfo {author} {\bibfnamefont {J.}~\bibnamefont {Wyman}},\ and\
  \bibinfo {author} {\bibfnamefont {J.-P.}\ \bibnamefont {Changeux}},\
  }\bibfield  {title} {\bibinfo {title} {On the nature of allosteric
  transitions: {A} plausible model},\ }\href
  {https://doi.org/10.1016/S0022-2836(65)80285-6} {\bibfield  {journal}
  {\bibinfo  {journal} {Journal of Molecular Biology}\ }\textbf {\bibinfo
  {volume} {12}},\ \bibinfo {pages} {88} (\bibinfo {year} {1965})}\BibitemShut
  {NoStop}%
\bibitem [{\citenamefont {Gillespie}(2007)}]{gillespie_stochastic_2007}%
  \BibitemOpen
  \bibfield  {author} {\bibinfo {author} {\bibfnamefont {D.~T.}\ \bibnamefont
  {Gillespie}},\ }\bibfield  {title} {\bibinfo {title} {Stochastic {Simulation}
  of {Chemical} {Kinetics}},\ }\href
  {https://doi.org/10.1146/annurev.physchem.58.032806.104637} {\bibfield
  {journal} {\bibinfo  {journal} {Annual Review of Physical Chemistry}\
  }\textbf {\bibinfo {volume} {58}},\ \bibinfo {pages} {35} (\bibinfo {year}
  {2007})}\BibitemShut {NoStop}%
\bibitem [{\citenamefont {Zakine}\ and\ \citenamefont
  {Vanden-Eijnden}(2023)}]{zakine_minimum-action_2023}%
  \BibitemOpen
  \bibfield  {author} {\bibinfo {author} {\bibfnamefont {R.}~\bibnamefont
  {Zakine}}\ and\ \bibinfo {author} {\bibfnamefont {E.}~\bibnamefont
  {Vanden-Eijnden}},\ }\bibfield  {title} {\bibinfo {title} {Minimum-{Action}
  {Method} for {Nonequilibrium} {Phase} {Transitions}},\ }\href
  {https://doi.org/10.1103/PhysRevX.13.041044} {\bibfield  {journal} {\bibinfo
  {journal} {Physical Review X}\ }\textbf {\bibinfo {volume} {13}},\ \bibinfo
  {pages} {041044} (\bibinfo {year} {2023})}\BibitemShut {NoStop}%
\bibitem [{\citenamefont {Hathcock}\ \emph {et~al.}(2024)\citenamefont
  {Hathcock}, \citenamefont {Yu},\ and\ \citenamefont
  {Tu}}]{hathcock_time-reversal_2024}%
  \BibitemOpen
  \bibfield  {author} {\bibinfo {author} {\bibfnamefont {D.}~\bibnamefont
  {Hathcock}}, \bibinfo {author} {\bibfnamefont {Q.}~\bibnamefont {Yu}},\ and\
  \bibinfo {author} {\bibfnamefont {Y.}~\bibnamefont {Tu}},\ }\bibfield
  {title} {\bibinfo {title} {Time-reversal symmetry breaking in the
  chemosensory array reveals a general mechanism for dissipation-enhanced
  cooperative sensing},\ }\href {https://doi.org/10.1038/s41467-024-52799-0}
  {\bibfield  {journal} {\bibinfo  {journal} {Nature Communications}\ }\textbf
  {\bibinfo {volume} {15}},\ \bibinfo {pages} {8892} (\bibinfo {year}
  {2024})}\BibitemShut {NoStop}%
\bibitem [{\citenamefont {Hill}(1910)}]{hill_possible_1910}%
  \BibitemOpen
  \bibfield  {author} {\bibinfo {author} {\bibfnamefont {A.~V.}\ \bibnamefont
  {Hill}},\ }\bibfield  {title} {\bibinfo {title} {The possible effects of the
  aggregation of the molecules of haemoglobin on its dissociation curves},\
  }\href {https://cir.nii.ac.jp/crid/1570291224727481088} {\bibfield  {journal}
  {\bibinfo  {journal} {J Physiol}\ }\textbf {\bibinfo {volume} {40}},\
  \bibinfo {pages} {4} (\bibinfo {year} {1910})}\BibitemShut {NoStop}%
\bibitem [{\citenamefont {Kullback}\ and\ \citenamefont
  {Leibler}(1951)}]{kullback_information_1951}%
  \BibitemOpen
  \bibfield  {author} {\bibinfo {author} {\bibfnamefont {S.}~\bibnamefont
  {Kullback}}\ and\ \bibinfo {author} {\bibfnamefont {R.~A.}\ \bibnamefont
  {Leibler}},\ }\bibfield  {title} {\bibinfo {title} {On {Information} and
  {Sufficiency}},\ }\href {https://doi.org/10.1214/aoms/1177729694} {\bibfield
  {journal} {\bibinfo  {journal} {The Annals of Mathematical Statistics}\
  }\textbf {\bibinfo {volume} {22}},\ \bibinfo {pages} {79} (\bibinfo {year}
  {1951})}\BibitemShut {NoStop}%
\bibitem [{\citenamefont {Kawai}\ \emph {et~al.}(2007)\citenamefont {Kawai},
  \citenamefont {Parrondo},\ and\ \citenamefont {den
  Broeck}}]{kawai_dissipation_2007}%
  \BibitemOpen
  \bibfield  {author} {\bibinfo {author} {\bibfnamefont {R.}~\bibnamefont
  {Kawai}}, \bibinfo {author} {\bibfnamefont {J.~M.~R.}\ \bibnamefont
  {Parrondo}},\ and\ \bibinfo {author} {\bibfnamefont {C.~V.}\ \bibnamefont
  {den Broeck}},\ }\bibfield  {title} {\bibinfo {title} {Dissipation: {The}
  {Phase}-{Space} {Perspective}},\ }\href
  {https://doi.org/10.1103/PhysRevLett.98.080602} {\bibfield  {journal}
  {\bibinfo  {journal} {Physical Review Letters}\ }\textbf {\bibinfo {volume}
  {98}},\ \bibinfo {pages} {080602} (\bibinfo {year} {2007})}\BibitemShut
  {NoStop}%
\bibitem [{\citenamefont {Yu}\ and\ \citenamefont {Tu}(2022)}]{yu_energy_2022}%
  \BibitemOpen
  \bibfield  {author} {\bibinfo {author} {\bibfnamefont {Q.}~\bibnamefont
  {Yu}}\ and\ \bibinfo {author} {\bibfnamefont {Y.}~\bibnamefont {Tu}},\
  }\bibfield  {title} {\bibinfo {title} {Energy {Cost} for {Flocking} of
  {Active} {Spins}: {The} {Cusped} {Dissipation} {Maximum} at the {Flocking}
  {Transition}},\ }\href {https://doi.org/10.1103/PhysRevLett.129.278001}
  {\bibfield  {journal} {\bibinfo  {journal} {Physical Review Letters}\
  }\textbf {\bibinfo {volume} {129}},\ \bibinfo {pages} {278001} (\bibinfo
  {year} {2022})}\BibitemShut {NoStop}%
\bibitem [{\citenamefont {Owen}\ and\ \citenamefont
  {Horowitz}(2023)}]{owen_size_2023}%
  \BibitemOpen
  \bibfield  {author} {\bibinfo {author} {\bibfnamefont {J.~A.}\ \bibnamefont
  {Owen}}\ and\ \bibinfo {author} {\bibfnamefont {J.~M.}\ \bibnamefont
  {Horowitz}},\ }\bibfield  {title} {\bibinfo {title} {Size limits the
  sensitivity of kinetic schemes},\ }\href
  {https://doi.org/10.1038/s41467-023-36705-8} {\bibfield  {journal} {\bibinfo
  {journal} {Nature Communications}\ }\textbf {\bibinfo {volume} {14}},\
  \bibinfo {pages} {1280} (\bibinfo {year} {2023})}\BibitemShut {NoStop}%
\bibitem [{\citenamefont {Lan}\ \emph {et~al.}(2012)\citenamefont {Lan},
  \citenamefont {Sartori}, \citenamefont {Neumann}, \citenamefont {Sourjik},\
  and\ \citenamefont {Tu}}]{lan_energyspeedaccuracy_2012}%
  \BibitemOpen
  \bibfield  {author} {\bibinfo {author} {\bibfnamefont {G.}~\bibnamefont
  {Lan}}, \bibinfo {author} {\bibfnamefont {P.}~\bibnamefont {Sartori}},
  \bibinfo {author} {\bibfnamefont {S.}~\bibnamefont {Neumann}}, \bibinfo
  {author} {\bibfnamefont {V.}~\bibnamefont {Sourjik}},\ and\ \bibinfo {author}
  {\bibfnamefont {Y.}~\bibnamefont {Tu}},\ }\bibfield  {title} {\bibinfo
  {title} {The energy–speed–accuracy trade-off in sensory adaptation},\
  }\href {https://doi.org/10.1038/nphys2276} {\bibfield  {journal} {\bibinfo
  {journal} {Nature Physics}\ }\textbf {\bibinfo {volume} {8}},\ \bibinfo
  {pages} {422} (\bibinfo {year} {2012})}\BibitemShut {NoStop}%
\bibitem [{\citenamefont {Sartori}\ and\ \citenamefont
  {Tu}(2015)}]{sartori_free_2015}%
  \BibitemOpen
  \bibfield  {author} {\bibinfo {author} {\bibfnamefont {P.}~\bibnamefont
  {Sartori}}\ and\ \bibinfo {author} {\bibfnamefont {Y.}~\bibnamefont {Tu}},\
  }\bibfield  {title} {\bibinfo {title} {Free {Energy} {Cost} of {Reducing}
  {Noise} while {Maintaining} a {High} {Sensitivity}},\ }\href
  {https://doi.org/10.1103/PhysRevLett.115.118102} {\bibfield  {journal}
  {\bibinfo  {journal} {Physical Review Letters}\ }\textbf {\bibinfo {volume}
  {115}},\ \bibinfo {pages} {118102} (\bibinfo {year} {2015})}\BibitemShut
  {NoStop}%
\bibitem [{\citenamefont {Fei}\ \emph {et~al.}(2018)\citenamefont {Fei},
  \citenamefont {Cao}, \citenamefont {Ouyang},\ and\ \citenamefont
  {Tu}}]{fei_design_2018}%
  \BibitemOpen
  \bibfield  {author} {\bibinfo {author} {\bibfnamefont {C.}~\bibnamefont
  {Fei}}, \bibinfo {author} {\bibfnamefont {Y.}~\bibnamefont {Cao}}, \bibinfo
  {author} {\bibfnamefont {Q.}~\bibnamefont {Ouyang}},\ and\ \bibinfo {author}
  {\bibfnamefont {Y.}~\bibnamefont {Tu}},\ }\bibfield  {title} {\bibinfo
  {title} {Design principles for enhancing phase sensitivity and suppressing
  phase fluctuations simultaneously in biochemical oscillatory systems},\
  }\href {https://doi.org/10.1038/s41467-018-03826-4} {\bibfield  {journal}
  {\bibinfo  {journal} {Nature Communications}\ }\textbf {\bibinfo {volume}
  {9}},\ \bibinfo {pages} {1434} (\bibinfo {year} {2018})}\BibitemShut
  {NoStop}%
\bibitem [{\citenamefont {Singh}\ \emph {et~al.}(2024)\citenamefont {Singh},
  \citenamefont {Sharma}, \citenamefont {Afanzar}, \citenamefont {Goldfarb},
  \citenamefont {Maklashina}, \citenamefont {Eisenbach}, \citenamefont
  {Cecchini},\ and\ \citenamefont {Iverson}}]{singh_cryoem_2024}%
  \BibitemOpen
  \bibfield  {author} {\bibinfo {author} {\bibfnamefont {P.~K.}\ \bibnamefont
  {Singh}}, \bibinfo {author} {\bibfnamefont {P.}~\bibnamefont {Sharma}},
  \bibinfo {author} {\bibfnamefont {O.}~\bibnamefont {Afanzar}}, \bibinfo
  {author} {\bibfnamefont {M.~H.}\ \bibnamefont {Goldfarb}}, \bibinfo {author}
  {\bibfnamefont {E.}~\bibnamefont {Maklashina}}, \bibinfo {author}
  {\bibfnamefont {M.}~\bibnamefont {Eisenbach}}, \bibinfo {author}
  {\bibfnamefont {G.}~\bibnamefont {Cecchini}},\ and\ \bibinfo {author}
  {\bibfnamefont {T.~M.}\ \bibnamefont {Iverson}},\ }\bibfield  {title}
  {\bibinfo {title} {{CryoEM} structures reveal how the bacterial flagellum
  rotates and switches direction},\ }\href
  {https://doi.org/10.1038/s41564-024-01674-1} {\bibfield  {journal} {\bibinfo
  {journal} {Nature Microbiology}\ }\textbf {\bibinfo {volume} {9}},\ \bibinfo
  {pages} {1271} (\bibinfo {year} {2024})}\BibitemShut {NoStop}%
\bibitem [{\citenamefont {Yuan}\ \emph {et~al.}(2010)\citenamefont {Yuan},
  \citenamefont {Fahrner}, \citenamefont {Turner},\ and\ \citenamefont
  {Berg}}]{yuan_asymmetry_2010}%
  \BibitemOpen
  \bibfield  {author} {\bibinfo {author} {\bibfnamefont {J.}~\bibnamefont
  {Yuan}}, \bibinfo {author} {\bibfnamefont {K.~A.}\ \bibnamefont {Fahrner}},
  \bibinfo {author} {\bibfnamefont {L.}~\bibnamefont {Turner}},\ and\ \bibinfo
  {author} {\bibfnamefont {H.~C.}\ \bibnamefont {Berg}},\ }\bibfield  {title}
  {\bibinfo {title} {Asymmetry in the clockwise and counterclockwise rotation
  of the bacterial flagellar motor},\ }\href
  {https://doi.org/10.1073/pnas.1007333107} {\bibfield  {journal} {\bibinfo
  {journal} {Proceedings of the National Academy of Sciences}\ }\textbf
  {\bibinfo {volume} {107}},\ \bibinfo {pages} {12846} (\bibinfo {year}
  {2010})}\BibitemShut {NoStop}%
\bibitem [{\citenamefont {van Albada}\ \emph {et~al.}(2009)\citenamefont {van
  Albada}, \citenamefont {Tănase‐Nicola},\ and\ \citenamefont {ten
  Wolde}}]{van_albada_switching_2009}%
  \BibitemOpen
  \bibfield  {author} {\bibinfo {author} {\bibfnamefont {S.~B.}\ \bibnamefont
  {van Albada}}, \bibinfo {author} {\bibfnamefont {S.}~\bibnamefont
  {Tănase‐Nicola}},\ and\ \bibinfo {author} {\bibfnamefont {P.~R.}\
  \bibnamefont {ten Wolde}},\ }\bibfield  {title} {\bibinfo {title} {The
  switching dynamics of the bacterial flagellar motor},\ }\href
  {https://doi.org/10.1038/msb.2009.74} {\bibfield  {journal} {\bibinfo
  {journal} {Molecular Systems Biology}\ }\textbf {\bibinfo {volume} {5}},\
  \bibinfo {pages} {316} (\bibinfo {year} {2009})}\BibitemShut {NoStop}%
\bibitem [{\citenamefont {Mora}\ \emph {et~al.}(2009)\citenamefont {Mora},
  \citenamefont {Yu},\ and\ \citenamefont {Wingreen}}]{mora_modeling_2009}%
  \BibitemOpen
  \bibfield  {author} {\bibinfo {author} {\bibfnamefont {T.}~\bibnamefont
  {Mora}}, \bibinfo {author} {\bibfnamefont {H.}~\bibnamefont {Yu}},\ and\
  \bibinfo {author} {\bibfnamefont {N.~S.}\ \bibnamefont {Wingreen}},\
  }\bibfield  {title} {\bibinfo {title} {Modeling {Torque} {Versus} {Speed},
  {Shot} {Noise}, and {Rotational} {Diffusion} of the {Bacterial} {Flagellar}
  {Motor}},\ }\href {https://doi.org/10.1103/PhysRevLett.103.248102} {\bibfield
   {journal} {\bibinfo  {journal} {Physical Review Letters}\ }\textbf {\bibinfo
  {volume} {103}},\ \bibinfo {pages} {248102} (\bibinfo {year}
  {2009})}\BibitemShut {NoStop}%
\bibitem [{\citenamefont {Meacci}\ and\ \citenamefont
  {Tu}(2009)}]{meacci_dynamics_2009}%
  \BibitemOpen
  \bibfield  {author} {\bibinfo {author} {\bibfnamefont {G.}~\bibnamefont
  {Meacci}}\ and\ \bibinfo {author} {\bibfnamefont {Y.}~\bibnamefont {Tu}},\
  }\bibfield  {title} {\bibinfo {title} {Dynamics of the bacterial flagellar
  motor with multiple stators},\ }\href
  {https://doi.org/10.1073/pnas.0809929106} {\bibfield  {journal} {\bibinfo
  {journal} {Proceedings of the National Academy of Sciences}\ }\textbf
  {\bibinfo {volume} {106}},\ \bibinfo {pages} {3746} (\bibinfo {year}
  {2009})}\BibitemShut {NoStop}%
\bibitem [{\citenamefont {Meacci}\ \emph {et~al.}(2011)\citenamefont {Meacci},
  \citenamefont {Lan},\ and\ \citenamefont {Tu}}]{meacci_dynamics_2011}%
  \BibitemOpen
  \bibfield  {author} {\bibinfo {author} {\bibfnamefont {G.}~\bibnamefont
  {Meacci}}, \bibinfo {author} {\bibfnamefont {G.}~\bibnamefont {Lan}},\ and\
  \bibinfo {author} {\bibfnamefont {Y.}~\bibnamefont {Tu}},\ }\bibfield
  {title} {\bibinfo {title} {Dynamics of the {Bacterial} {Flagellar} {Motor}:
  {The} {Effects} of {Stator} {Compliance}, {Back} {Steps}, {Temperature}, and
  {Rotational} {Asymmetry}},\ }\href
  {https://doi.org/10.1016/j.bpj.2011.02.045} {\bibfield  {journal} {\bibinfo
  {journal} {Biophysical Journal}\ }\textbf {\bibinfo {volume} {100}},\
  \bibinfo {pages} {1986} (\bibinfo {year} {2011})}\BibitemShut {NoStop}%
\bibitem [{\citenamefont {Tu}\ and\ \citenamefont
  {Cao}(2018)}]{tu_design_2018}%
  \BibitemOpen
  \bibfield  {author} {\bibinfo {author} {\bibfnamefont {Y.}~\bibnamefont
  {Tu}}\ and\ \bibinfo {author} {\bibfnamefont {Y.}~\bibnamefont {Cao}},\
  }\bibfield  {title} {\bibinfo {title} {Design principles and optimal
  performance for molecular motors under realistic constraints},\ }\href
  {https://doi.org/10.1103/PhysRevE.97.022403} {\bibfield  {journal} {\bibinfo
  {journal} {Physical Review E}\ }\textbf {\bibinfo {volume} {97}},\ \bibinfo
  {pages} {022403} (\bibinfo {year} {2018})}\BibitemShut {NoStop}%
\bibitem [{\citenamefont {Cao}\ \emph {et~al.}(2022)\citenamefont {Cao},
  \citenamefont {Li},\ and\ \citenamefont {Tu}}]{cao_modeling_2022}%
  \BibitemOpen
  \bibfield  {author} {\bibinfo {author} {\bibfnamefont {Y.}~\bibnamefont
  {Cao}}, \bibinfo {author} {\bibfnamefont {T.}~\bibnamefont {Li}},\ and\
  \bibinfo {author} {\bibfnamefont {Y.}~\bibnamefont {Tu}},\ }\bibfield
  {title} {\bibinfo {title} {Modeling {Bacterial} {Flagellar} {Motor} {With}
  {New} {Structure} {Information}: {Rotational} {Dynamics} of {Two}
  {Interacting} {Protein} {Nano}-{Rings}},\ }\href
  {https://www.frontiersin.org/articles/10.3389/fmicb.2022.866141} {\bibfield
  {journal} {\bibinfo  {journal} {Frontiers in Microbiology}\ }\textbf
  {\bibinfo {volume} {13}} (\bibinfo {year} {2022})}\BibitemShut {NoStop}%
\bibitem [{\citenamefont {Delalez}\ \emph {et~al.}(2010)\citenamefont
  {Delalez}, \citenamefont {Wadhams}, \citenamefont {Rosser}, \citenamefont
  {Xue}, \citenamefont {Brown}, \citenamefont {Dobbie}, \citenamefont {Berry},
  \citenamefont {Leake},\ and\ \citenamefont
  {Armitage}}]{delalez_signal-dependent_2010}%
  \BibitemOpen
  \bibfield  {author} {\bibinfo {author} {\bibfnamefont {N.~J.}\ \bibnamefont
  {Delalez}}, \bibinfo {author} {\bibfnamefont {G.~H.}\ \bibnamefont
  {Wadhams}}, \bibinfo {author} {\bibfnamefont {G.}~\bibnamefont {Rosser}},
  \bibinfo {author} {\bibfnamefont {Q.}~\bibnamefont {Xue}}, \bibinfo {author}
  {\bibfnamefont {M.~T.}\ \bibnamefont {Brown}}, \bibinfo {author}
  {\bibfnamefont {I.~M.}\ \bibnamefont {Dobbie}}, \bibinfo {author}
  {\bibfnamefont {R.~M.}\ \bibnamefont {Berry}}, \bibinfo {author}
  {\bibfnamefont {M.~C.}\ \bibnamefont {Leake}},\ and\ \bibinfo {author}
  {\bibfnamefont {J.~P.}\ \bibnamefont {Armitage}},\ }\bibfield  {title}
  {\bibinfo {title} {Signal-dependent turnover of the bacterial flagellar
  switch protein {FliM}},\ }\href {https://doi.org/10.1073/pnas.1000284107}
  {\bibfield  {journal} {\bibinfo  {journal} {Proceedings of the National
  Academy of Sciences}\ }\textbf {\bibinfo {volume} {107}},\ \bibinfo {pages}
  {11347} (\bibinfo {year} {2010})}\BibitemShut {NoStop}%
\bibitem [{\citenamefont {Yuan}\ \emph {et~al.}(2012)\citenamefont {Yuan},
  \citenamefont {Branch}, \citenamefont {Hosu},\ and\ \citenamefont
  {Berg}}]{yuan_adaptation_2012}%
  \BibitemOpen
  \bibfield  {author} {\bibinfo {author} {\bibfnamefont {J.}~\bibnamefont
  {Yuan}}, \bibinfo {author} {\bibfnamefont {R.~W.}\ \bibnamefont {Branch}},
  \bibinfo {author} {\bibfnamefont {B.~G.}\ \bibnamefont {Hosu}},\ and\
  \bibinfo {author} {\bibfnamefont {H.~C.}\ \bibnamefont {Berg}},\ }\bibfield
  {title} {\bibinfo {title} {Adaptation at the output of the chemotaxis
  signalling pathway},\ }\href {https://doi.org/10.1038/nature10964} {\bibfield
   {journal} {\bibinfo  {journal} {Nature}\ }\textbf {\bibinfo {volume}
  {484}},\ \bibinfo {pages} {233} (\bibinfo {year} {2012})}\BibitemShut
  {NoStop}%
\bibitem [{\citenamefont {Lele}\ \emph {et~al.}(2012)\citenamefont {Lele},
  \citenamefont {Branch}, \citenamefont {Nathan},\ and\ \citenamefont
  {Berg}}]{lele_mechanism_2012}%
  \BibitemOpen
  \bibfield  {author} {\bibinfo {author} {\bibfnamefont {P.~P.}\ \bibnamefont
  {Lele}}, \bibinfo {author} {\bibfnamefont {R.~W.}\ \bibnamefont {Branch}},
  \bibinfo {author} {\bibfnamefont {V.~S.~J.}\ \bibnamefont {Nathan}},\ and\
  \bibinfo {author} {\bibfnamefont {H.~C.}\ \bibnamefont {Berg}},\ }\bibfield
  {title} {\bibinfo {title} {Mechanism for adaptive remodeling of the bacterial
  flagellar switch},\ }\href {https://doi.org/10.1073/pnas.1212327109}
  {\bibfield  {journal} {\bibinfo  {journal} {Proceedings of the National
  Academy of Sciences}\ }\textbf {\bibinfo {volume} {109}},\ \bibinfo {pages}
  {20018} (\bibinfo {year} {2012})}\BibitemShut {NoStop}%
\bibitem [{\citenamefont {Delalez}\ \emph {et~al.}(2014)\citenamefont
  {Delalez}, \citenamefont {Berry},\ and\ \citenamefont
  {Armitage}}]{delalez_stoichiometry_2014}%
  \BibitemOpen
  \bibfield  {author} {\bibinfo {author} {\bibfnamefont {N.~J.}\ \bibnamefont
  {Delalez}}, \bibinfo {author} {\bibfnamefont {R.~M.}\ \bibnamefont {Berry}},\
  and\ \bibinfo {author} {\bibfnamefont {J.~P.}\ \bibnamefont {Armitage}},\
  }\bibfield  {title} {\bibinfo {title} {Stoichiometry and {Turnover} of the
  {Bacterial} {Flagellar} {Switch} {Protein} {FliN}},\ }\href
  {https://doi.org/10.1128/mbio.01216-14} {\bibfield  {journal} {\bibinfo
  {journal} {mBio}\ }\textbf {\bibinfo {volume} {5}},\ \bibinfo {pages}
  {10.1128/mbio.01216} (\bibinfo {year} {2014})}\BibitemShut {NoStop}%
\bibitem [{\citenamefont {Branch}\ \emph {et~al.}(2014)\citenamefont {Branch},
  \citenamefont {Sayegh}, \citenamefont {Shen}, \citenamefont {Nathan},\ and\
  \citenamefont {Berg}}]{branch_adaptive_2014}%
  \BibitemOpen
  \bibfield  {author} {\bibinfo {author} {\bibfnamefont {R.~W.}\ \bibnamefont
  {Branch}}, \bibinfo {author} {\bibfnamefont {M.~N.}\ \bibnamefont {Sayegh}},
  \bibinfo {author} {\bibfnamefont {C.}~\bibnamefont {Shen}}, \bibinfo {author}
  {\bibfnamefont {V.~S.~J.}\ \bibnamefont {Nathan}},\ and\ \bibinfo {author}
  {\bibfnamefont {H.~C.}\ \bibnamefont {Berg}},\ }\bibfield  {title} {\bibinfo
  {title} {Adaptive {Remodelling} by {FliN} in the {Bacterial} {Rotary}
  {Motor}},\ }\href {https://doi.org/10.1016/j.jmb.2014.07.009} {\bibfield
  {journal} {\bibinfo  {journal} {Journal of Molecular Biology}\ }\textbf
  {\bibinfo {volume} {426}},\ \bibinfo {pages} {3314} (\bibinfo {year}
  {2014})}\BibitemShut {NoStop}%
\bibitem [{\citenamefont {Antani}\ \emph {et~al.}(2021)\citenamefont {Antani},
  \citenamefont {Gupta}, \citenamefont {Lee}, \citenamefont {Rhee},
  \citenamefont {Manson},\ and\ \citenamefont
  {Lele}}]{antani_mechanosensitive_2021}%
  \BibitemOpen
  \bibfield  {author} {\bibinfo {author} {\bibfnamefont {J.~D.}\ \bibnamefont
  {Antani}}, \bibinfo {author} {\bibfnamefont {R.}~\bibnamefont {Gupta}},
  \bibinfo {author} {\bibfnamefont {A.~H.}\ \bibnamefont {Lee}}, \bibinfo
  {author} {\bibfnamefont {K.~Y.}\ \bibnamefont {Rhee}}, \bibinfo {author}
  {\bibfnamefont {M.~D.}\ \bibnamefont {Manson}},\ and\ \bibinfo {author}
  {\bibfnamefont {P.~P.}\ \bibnamefont {Lele}},\ }\bibfield  {title} {\bibinfo
  {title} {Mechanosensitive recruitment of stator units promotes binding of the
  response regulator {CheY}-{P} to the flagellar motor},\ }\href
  {https://doi.org/10.1038/s41467-021-25774-2} {\bibfield  {journal} {\bibinfo
  {journal} {Nature Communications}\ }\textbf {\bibinfo {volume} {12}},\
  \bibinfo {pages} {5442} (\bibinfo {year} {2021})}\BibitemShut {NoStop}%
\bibitem [{\citenamefont {Gross}(2004)}]{gross_hither_2004}%
  \BibitemOpen
  \bibfield  {author} {\bibinfo {author} {\bibfnamefont {S.~P.}\ \bibnamefont
  {Gross}},\ }\bibfield  {title} {\bibinfo {title} {Hither and yon: a review of
  bi-directional microtubule-based transport},\ }\href
  {https://doi.org/10.1088/1478-3967/1/2/R01} {\bibfield  {journal} {\bibinfo
  {journal} {Physical Biology}\ }\textbf {\bibinfo {volume} {1}},\ \bibinfo
  {pages} {R1} (\bibinfo {year} {2004})}\BibitemShut {NoStop}%
\bibitem [{\citenamefont {Welte}(2004)}]{welte_bidirectional_2004}%
  \BibitemOpen
  \bibfield  {author} {\bibinfo {author} {\bibfnamefont {M.~A.}\ \bibnamefont
  {Welte}},\ }\bibfield  {title} {\bibinfo {title} {Bidirectional {Transport}
  along {Microtubules}},\ }\href {https://doi.org/10.1016/j.cub.2004.06.045}
  {\bibfield  {journal} {\bibinfo  {journal} {Current Biology}\ }\textbf
  {\bibinfo {volume} {14}},\ \bibinfo {pages} {R525} (\bibinfo {year}
  {2004})}\BibitemShut {NoStop}%
\bibitem [{\citenamefont {Hancock}(2014)}]{hancock_bidirectional_2014}%
  \BibitemOpen
  \bibfield  {author} {\bibinfo {author} {\bibfnamefont {W.~O.}\ \bibnamefont
  {Hancock}},\ }\bibfield  {title} {\bibinfo {title} {Bidirectional cargo
  transport: moving beyond tug of war},\ }\href
  {https://doi.org/10.1038/nrm3853} {\bibfield  {journal} {\bibinfo  {journal}
  {Nature Reviews Molecular Cell Biology}\ }\textbf {\bibinfo {volume} {15}},\
  \bibinfo {pages} {615} (\bibinfo {year} {2014})}\BibitemShut {NoStop}%
\bibitem [{\citenamefont {Müller}\ \emph {et~al.}(2008)\citenamefont
  {Müller}, \citenamefont {Klumpp},\ and\ \citenamefont
  {Lipowsky}}]{muller_tug--war_2008}%
  \BibitemOpen
  \bibfield  {author} {\bibinfo {author} {\bibfnamefont {M.~J.~I.}\
  \bibnamefont {Müller}}, \bibinfo {author} {\bibfnamefont {S.}~\bibnamefont
  {Klumpp}},\ and\ \bibinfo {author} {\bibfnamefont {R.}~\bibnamefont
  {Lipowsky}},\ }\bibfield  {title} {\bibinfo {title} {Tug-of-war as a
  cooperative mechanism for bidirectional cargo transport by molecular
  motors},\ }\href {https://doi.org/10.1073/pnas.0706825105} {\bibfield
  {journal} {\bibinfo  {journal} {Proceedings of the National Academy of
  Sciences}\ }\textbf {\bibinfo {volume} {105}},\ \bibinfo {pages} {4609}
  (\bibinfo {year} {2008})}\BibitemShut {NoStop}%
\bibitem [{\citenamefont {Soppina}\ \emph {et~al.}(2009)\citenamefont
  {Soppina}, \citenamefont {Rai}, \citenamefont {Ramaiya}, \citenamefont
  {Barak},\ and\ \citenamefont {Mallik}}]{soppina_tug--war_2009}%
  \BibitemOpen
  \bibfield  {author} {\bibinfo {author} {\bibfnamefont {V.}~\bibnamefont
  {Soppina}}, \bibinfo {author} {\bibfnamefont {A.~K.}\ \bibnamefont {Rai}},
  \bibinfo {author} {\bibfnamefont {A.~J.}\ \bibnamefont {Ramaiya}}, \bibinfo
  {author} {\bibfnamefont {P.}~\bibnamefont {Barak}},\ and\ \bibinfo {author}
  {\bibfnamefont {R.}~\bibnamefont {Mallik}},\ }\bibfield  {title} {\bibinfo
  {title} {Tug-of-war between dissimilar teams of microtubule motors regulates
  transport and fission of endosomes},\ }\href
  {https://doi.org/10.1073/pnas.0906524106} {\bibfield  {journal} {\bibinfo
  {journal} {Proceedings of the National Academy of Sciences}\ }\textbf
  {\bibinfo {volume} {106}},\ \bibinfo {pages} {19381} (\bibinfo {year}
  {2009})}\BibitemShut {NoStop}%
\bibitem [{\citenamefont {Müller}\ \emph {et~al.}(2010)\citenamefont
  {Müller}, \citenamefont {Klumpp},\ and\ \citenamefont
  {Lipowsky}}]{muller_bidirectional_2010}%
  \BibitemOpen
  \bibfield  {author} {\bibinfo {author} {\bibfnamefont {M.~J.~I.}\
  \bibnamefont {Müller}}, \bibinfo {author} {\bibfnamefont {S.}~\bibnamefont
  {Klumpp}},\ and\ \bibinfo {author} {\bibfnamefont {R.}~\bibnamefont
  {Lipowsky}},\ }\bibfield  {title} {\bibinfo {title} {Bidirectional
  {Transport} by {Molecular} {Motors}: {Enhanced} {Processivity} and {Response}
  to {External} {Forces}},\ }\href {https://doi.org/10.1016/j.bpj.2010.02.037}
  {\bibfield  {journal} {\bibinfo  {journal} {Biophysical Journal}\ }\textbf
  {\bibinfo {volume} {98}},\ \bibinfo {pages} {2610} (\bibinfo {year}
  {2010})}\BibitemShut {NoStop}%
\bibitem [{\citenamefont {Kunwar}\ \emph {et~al.}(2011)\citenamefont {Kunwar},
  \citenamefont {Tripathy}, \citenamefont {Xu}, \citenamefont {Mattson},
  \citenamefont {Anand}, \citenamefont {Sigua}, \citenamefont {Vershinin},
  \citenamefont {McKenney}, \citenamefont {Yu}, \citenamefont {Mogilner},\ and\
  \citenamefont {Gross}}]{kunwar_mechanical_2011}%
  \BibitemOpen
  \bibfield  {author} {\bibinfo {author} {\bibfnamefont {A.}~\bibnamefont
  {Kunwar}}, \bibinfo {author} {\bibfnamefont {S.~K.}\ \bibnamefont
  {Tripathy}}, \bibinfo {author} {\bibfnamefont {J.}~\bibnamefont {Xu}},
  \bibinfo {author} {\bibfnamefont {M.~K.}\ \bibnamefont {Mattson}}, \bibinfo
  {author} {\bibfnamefont {P.}~\bibnamefont {Anand}}, \bibinfo {author}
  {\bibfnamefont {R.}~\bibnamefont {Sigua}}, \bibinfo {author} {\bibfnamefont
  {M.}~\bibnamefont {Vershinin}}, \bibinfo {author} {\bibfnamefont {R.~J.}\
  \bibnamefont {McKenney}}, \bibinfo {author} {\bibfnamefont {C.~C.}\
  \bibnamefont {Yu}}, \bibinfo {author} {\bibfnamefont {A.}~\bibnamefont
  {Mogilner}},\ and\ \bibinfo {author} {\bibfnamefont {S.~P.}\ \bibnamefont
  {Gross}},\ }\bibfield  {title} {\bibinfo {title} {Mechanical stochastic
  tug-of-war models cannot explain bidirectional lipid-droplet transport},\
  }\href {https://doi.org/10.1073/pnas.1107841108} {\bibfield  {journal}
  {\bibinfo  {journal} {Proceedings of the National Academy of Sciences}\
  }\textbf {\bibinfo {volume} {108}},\ \bibinfo {pages} {18960} (\bibinfo
  {year} {2011})}\BibitemShut {NoStop}%
\bibitem [{\citenamefont {D’Souza}\ \emph {et~al.}(2023)\citenamefont
  {D’Souza}, \citenamefont {Grover}, \citenamefont {Monzon}, \citenamefont
  {Santen},\ and\ \citenamefont {Diez}}]{dsouza_vesicles_2023}%
  \BibitemOpen
  \bibfield  {author} {\bibinfo {author} {\bibfnamefont {A.~I.}\ \bibnamefont
  {D’Souza}}, \bibinfo {author} {\bibfnamefont {R.}~\bibnamefont {Grover}},
  \bibinfo {author} {\bibfnamefont {G.~A.}\ \bibnamefont {Monzon}}, \bibinfo
  {author} {\bibfnamefont {L.}~\bibnamefont {Santen}},\ and\ \bibinfo {author}
  {\bibfnamefont {S.}~\bibnamefont {Diez}},\ }\bibfield  {title} {\bibinfo
  {title} {Vesicles driven by dynein and kinesin exhibit directional reversals
  without regulators},\ }\href {https://doi.org/10.1038/s41467-023-42605-8}
  {\bibfield  {journal} {\bibinfo  {journal} {Nature Communications}\ }\textbf
  {\bibinfo {volume} {14}},\ \bibinfo {pages} {7532} (\bibinfo {year}
  {2023})}\BibitemShut {NoStop}%
\end{thebibliography}%

\end{document}